\documentclass[english,fleqn,epsfig,psfig,showkeys,calligra,lscape,12pt,aipproc,nohypertex]{revtex4-1}
\usepackage{epsfig}
\usepackage[T1]{fontenc}
\usepackage{german}
\usepackage[latin1]{inputenc}
\usepackage{graphics}
\usepackage{graphicx}
\usepackage{multirow}
\usepackage{float}
\usepackage{bm}
\usepackage{color}   
\usepackage{dcolumn} 
\usepackage{amsmath}
\usepackage{amssymb}
\usepackage{xcolor}
\usepackage{babel}
\usepackage[twoside]{rotating}
\begin{document}
\scriptsize
\title{Tunneling time in attosecond experiments and time-energy 
uncertainty relation} 
\author{Ossama Kullie} 
\affiliation{Theoretical Physics, Institute for Physics, Department 
of Mathematics and Natural Science, University of Kassel, Germany}
\thanks{\tiny Electronic mail: kullie@uni-kassel.de}
\begin{abstract} 
\scriptsize
In this work we present a theoretical model supported with a physical reasoning 
 leading to a relation which performs an excellent estimation for the tunneling 
time in attosecond and strong field experiments,  
where we address the important case of the He-atom \cite{Eckle:2008s,Eckle:2008}. 
Our tunneling time estimation is found by utilizing the time-energy uncertainty 
relation and represents a quantum clock. The tunneling time is also featured as 
the time of passage through the barrier similarly to the Einstein's 
{\it photon box Gedanken experiment}. 
Our work tackles an important study case for the theory of time in quantum 
mechanics, and is very promising for the search for a (general) time operator 
in quantum mechanics. 
The work can be seen as a new fundamental step in dealing with the tunneling 
time in strong field and ultra-fast science, and is appealing for more 
elaborate treatments using quantum wave packet dynamics and especially for 
complex atoms and molecules.                                                          
\end{abstract} 
\maketitle                                                                                                                                  
\section{Introduction}\label{sec:int}                                                                                                                         
A comprehensive theory of time measurement in quantum mechanics is missing to 
date \cite{Muga:2008} (chap. 3). Often is said that time plays a role essentially 
different from the role of the position in quantum mechanics.  
In contrast Hilgevoord \cite{Hilgevoord:2002} argued that there is nothing in the 
formalism of the quantum mechanics that forces us to treat time and position 
differently. Observables as position, velocity, etc. both in classical mechanics 
as well as in quantum mechanics, are relative observables, and  one never 
measures the absolute position of a particle, but the distance in between 
the particle and some other object \cite{Hilgevoord:2002,Aharonov:2000}. 
Indeed there is many attempts to consider time  as  a dynamical intrinsic, 
or  an observable time called event time. Hilgevoord concluded 
\cite{Hilgevoord:2002} that when looking to a time operator  a distinction 
must be made between universal time coordinate $t$, a c-number like a space 
coordinate, and the dynamical time variable of a physical system suited 
in space-time, i.e. clocks.   
Busch  \cite{Busch:1990I,Busch:1990II} argued that the conundrum of 
the time-energy uncertainty relation (TEUR)  in quantum mechanics is related 
in first place to the fact  that the time is identified as a parameter in 
Schr\"odinger equation (SEQ).
He classified three types of time in quantum mechanics: external time 
(parametric or laboratory time), intrinsic or dynamical time and observable time. 
External time are carried out with clocks that are not dynamically connected with 
the object studied in the experiment, and usually called parametric time. 
The intrinsic or dynamical time is measured in term of the physical system 
undergoing dynamically a change, where every  dynamical variable marks the passage 
of time, we will see this is important for our time invention where the energy 
serves as the dynamical variable in question, which enables a quantitative measure 
for the length of the time interval of the tunneling or the tunneling time (T-time) 
in strong field experiments. 
The third type according to Busch is the observable time  or event time, for 
example the time of arrival of the decay products at a detector. 

In the history of the quantum mechanics,  the earliest attempt,   
which  causes one of the most impressive  debates, 
is the  Einstein's {\it photon box Gedanken experiment} \cite{Cooke:1980} 
or the Bohr-Einstein weighing {\it photon box Gedanken experiment} 
\cite{Aharonov:2000} (and the references therein).
A photon is allowed to escape from a box through a hole, which is closed and 
opened temporarily  by a shutter, the period of time is determined by a clock, 
which is part of the box system, that means the time is intrinsic and 
dynamically connected with the system under consideration. 
 The total mass of the box before and after a photon passes is measured. 
Bohr showed that the process of weighting introduces a quantum uncertainty 
(in the gravitational field) leading to an uncertainty in time $\tau$, which 
is the time needed to pass out of the box that usually called the time of passage 
\cite{Busch:1990I,Busch:1990II}, in accordance with the TEUR, 
eq (\ref{ucr}) below. 
Aharonov and Rezinik \cite{Aharonov:2000} offer a similar interpretation, that
the weighing leads, due to the back reaction of the system underlying  
a perturbation (energy measurement), to an uncertainty in the time of the 
internal clock relative to  the external time \cite{Aharonov:2000}. 
Hence for quantum systems it is important to observe the time from within 
the system or using an internal clock. 
Busch \cite{Busch:1990I} presented an argument which makes no assumptions 
concerning the method of measurement, and simply based only on a version of 
quantum clock uncertainty relation as follows, if the energy of the escaping 
photons  is determined with an accuracy $\delta E$ from the difference of energy 
before and after the opening period of the shutter, then these energies must be 
defined within an uncertainty $\delta E$, i.e the box energy uncertainty 
$\Delta E$ must satisfy $\Delta E\le \delta E $. Then the clock uncertainty 
allows to conclude that the box system needs at least a time 
$t_0=\frac{\hbar}{\Delta E}$ in order to evolve from the initial state ``shutter 
closed'' to the  orthogonal state  ``shutter open''. Accordingly, the time 
interval within which a photon can pass the shutter is indeterminate by an 
amount $\Delta T=t_0$. 
This leads to Bohr's TEUR $\Delta T\delta E \approx \hbar$ 
\cite{Muga:2008} (chap. 3). 

In this work we use similar ideas, where the T-times under the barrier 
(denoted $\tau_{_{T,d}}$) is suggested to be similar to the time of passage 
through the barrier (and escaping at the exit of the barrier), and the (quantum) 
particle (an electron) undergoes this process spends a time that is the time 
needed from the moment of entering the barrier to the  moment of escaping from 
under the barrier in the tunneling direction. 
In addition we suggest a time interval needed to reach the entrance of the barrier 
(denoted $\tau_{_{T,i}}$), after it is shaken off by the laser field at its initial 
position $x_i$. 
$\tau_{_{T,d}}$ is similar to the traversal time used in context of the tunneling 
approaches \cite{Hauge:1989,Landauer:1994} or the Feynman path integral (FPI) 
approach \cite{Yamada:2004,Sokolovski:1994,Sokolovski:1990} 
(and \cite{Muga:2008} chap. 7).
But in contrast we do not make any assumption about paths inside the barrier, 
where as well-known, the FPI approach is based on all paths starting at the 
entrance of the barrier at $t=0$  and end at the exit of the barrier at time 
$t=\tau$, which defines a time duration $\tau$. 
A second type of T-time that we invent is what we call the symmetrical T-time or 
the total T-time (denoted $\tau_{_{T,sym}}$). We will see that can be easily 
calculated from the symmetry property of the T-time but then later we found
$\tau_{_{T,sym}}=\tau_{_{T,i}}+\tau_{_{T,d}}$,  that is 
the time is accounted from the moment of starting the interaction  process, where 
the electron getting a shake-off, responses and jumping up to the tunneling 
``entrance`` point taking the (opposite) orientation of the field, moving then 
under the barrier (the tunneling) to the ''exit'' point and escapes the barrier 
to the continuum.
The key issue of the present work is (in the words of Busch 
\cite{Muga:2008}) a case study, the T-time in attosecond experiment 
and ultra-fast science derived by utilizing the time-energy uncertainty 
relation (TEUR): 
\begin{equation}\label{ucr} 
  {\,\, \Delta T  \cdot \Delta E \ge \frac{\hbar}{2}}
\end{equation}
In sec \ref{sec:TM} we present our theoretical model,  
in sec \ref{sec:pr} we offer a convincing physical reasoning for our theoretical 
model, in \ref{sec:dis} we discuss our result with a comparison to the experiment 
and finally we give a conclusion to our work.                                          
\vspace*{-0.5cm}                                                      
\section{theory and model}\label{sec:TM}\vspace{-0.3cm}
\vspace*{-0.250cm}
\subsection{Preview}\label{ssec:prew}
\begin{figure}[t]
\vspace{-5.1cm}
\hspace*{-0.25cm}
\includegraphics[height=17.0cm,width=13.5cm]{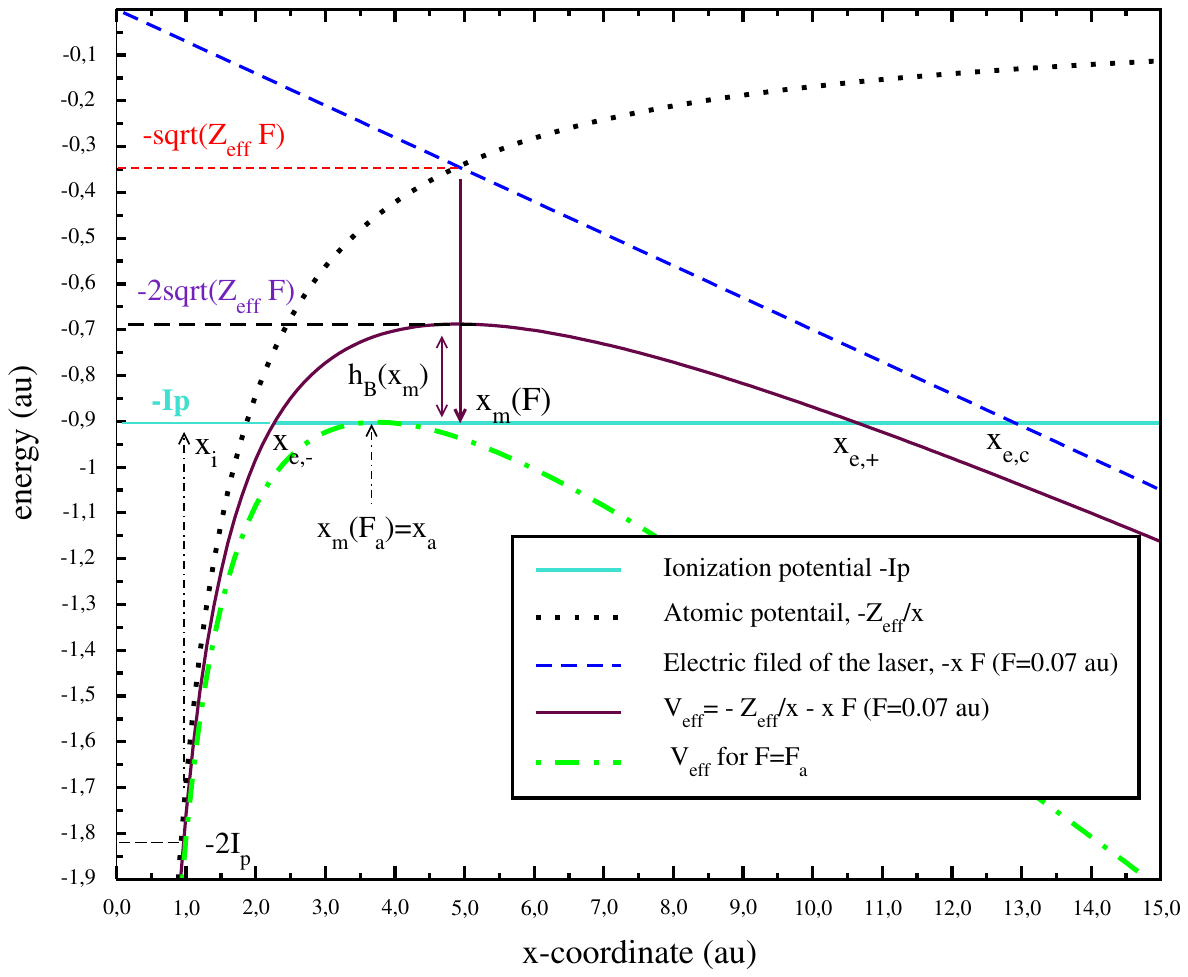}
\vspace{-6.50cm}
\caption{\label{fig:ptc}\tiny 
Graphic display of the potential curves, the barrier width and the two inner 
and outer points $x_{e,\pm}={(I_p\pm\delta_z)/2F}$, the ''classical exit`` point 
$x_{e,c}=I_p/F$ and the $x_m(F)=\sqrt{Z_{_{eff}/F}}$  the position 
at maximum of the barrier height, (note $x_a=x_m(F=F_a)$) see text.}
\vspace{-0.30cm}
\end{figure}

\vspace*{-0.250cm}
In the following we suggest a way to approximate the T-time in attosecond 
experiment based on simple mathematical and quantum mechanical rules. 
Our start is a model of Augst et al. \cite{Augst:1989,Augst:1991}, 
where the appearance intensity of a laser pulse for the ionization of the noble 
gases is predicted. 
The appearance intensity is defined \cite{Augst:1989} as the intensity at which 
a small number of ions is produced.  In this model (in atomic units) 
the effective potential of the atom-laser system is given  by
\begin{equation}\label{Vx}
V_{eff}(x)=V(x)-x F =-\frac{Z_{eff}}{x}-x F,
\end{equation}
where $F=F_m$ is the field strength at maximum of the laser pulse 
 (in this work in all our formulas $F$ stands for $F_m$), and $Z_{eff}$ is the
effective nuclear charge that can be found by treating the (active) electron 
orbital as hydrogen-like, similarly to the well-known single-active-electron  
(SAE) model \cite{Muller:1999,Tong:2005}.   
The choice of $Z_{eff}$ is easy recognized for many electron system and
 well-known in atomic, molecular  and  plasma physics 
\cite{Schlueter:1987,Lange:1992,Dreissigacker:2013}. 
 We take one dimensional model along the x-axis as justified by Klaiber and 
Yakaboylu et al \cite{Klaiber:2013,Yakaboylu:2013}.
Augst et al \cite{Augst:1989} calculated the position of the barrier maximum 
$x_{m}$  by setting 
$\partial{V_{eff}(x)}/\partial{x}=0 \Rightarrow x_a=x_m(F_a)=(\sqrt{Z_{eff}/F_a})$ 
and by equating  $V_{eff}(x_{m})$ to the ionization potential $V_{eff}(x_a)=-I_p$ 
(compare fig \ref{fig:ptc}, the lower green curve) they found an expression 
for the atomic field strength $F_a$, 
\begin{equation}\label{Fa}
-\frac{Z_{eff}}{x_{a}}-x_{a} F =-I_{p}
\Rightarrow F_a=\frac{I_{p}^{2}}{4 Z_{eff}}
\end{equation}
and the appearance intensity $I_a=F^2_a$. 
Now we take this idea and relate our argumentation to this model for $F\le F_a$, 
i.e. for the tunnel ionization in the regime of the well-known Keldysh 
\cite{Keldysh:1964} parameter $\gamma_K<1$. 
It is easy to see  (see fig \ref{fig:ptc}) that the tunnel exit obeys 
$x_{e,+}(F) \ge x_{m}(F)$ 
(note the equality is valid for $F=F_a$, and for the subscript $+$ see below) 
and the energy of the tunneling electron is not sufficient to reach the the top 
of the barrier (as for $F_a$) suggesting an energy uncertainty, which is 
determined by the energy the electron needs to overcome the barrier and 
appears in the continuum at the exit point.
It appears with zero velocity at the exit point $x_{e,+}$ 
according to the strong-field approximation (SFA) due to 
Keldysh-Faisal-Reiss theory \cite{Keldysh:1964,Faisal:1973,Reiss:1980}.
Indeed the barrier height at a position $x$ is, compare fig \ref{fig:ptc}:
\begin{equation}\label{hbx}
\overline{h_B(x)}=\mid h_B(x)\mid=\mid E-V_{_{eff}}(x)\mid=
\mid- I_p+\frac{Z_{eff}}{x}+x F\mid
\end{equation}
that is equal to the difference between the ionization potential and 
effective potential $V_{eff}(x)$ of the system (atom+laser) at the position $x$, 
where $E=-I_p$ is the binding energy of the electron before interacting 
with the laser. Note we can also get $x_m$ and the maximum $h_B(x_m)$ from the 
derivative of eq (\ref{hbx}), ${\partial h_B(x)}/{\partial x}=0$.   
An immediately arising question is, what about its maximum $h_B(x_m)$
and the energy uncertainty when the electron passes the barrier and is shifted 
to the continuum or a ``quasi`` energy level?.
First in regard to the derivation  of $x_a=x_m(F_a)$ and the eq (\ref{Fa}) of Augst 
et al,  it turns out (compare fig \ref{fig:ptc}) that the maximum of the barrier 
height $h_B(x_m)$ for arbitrary filed strength lies at  
$\, x_m(F)=\sqrt{Z_{eff}/F}$. 
This follows immediately from the fact that $x_m$ is determined by the maximum 
of the effective potential energy for arbitrary filed strength and that is  
the intersection point of the two potentials, $V(x)=-\frac{Zeff}{x}$ and $-x\,F$ 
(see fig \ref{fig:ptc}), then
\begin{equation}\label{xm}
\frac{-Z_{eff}}{x}=-x\,F\Rightarrow x_m(F)=\sqrt{\frac{Z_{eff}}{F}}
\end{equation}  
Otherwise (to the both sides) one of the two potentials 
($-\frac{Z_{eff}}{x} \mbox{ or } -x F$) slopes down stronger than the other 
 slops up (which can be easily gathered from fig \ref{fig:ptc} and eq (\ref{xm})), 
leading to $V_{_{eff}}(x)<V_{_{eff}}(x_m)$ for $x\ne x_m$. 
Indeed eq (\ref{Fa}) can be generalized as the following, for a filed strength
$F\le F_a$  we get
\begin{equation}\label{FFa}
F\le\frac{I_{p}^{2}}{4 Z_{eff}}\Rightarrow \delta_z^2= I_p^2-4 Z_{eff} F\ge 0
\end{equation}
The equality $\delta_z=0$ is valid for $F=F_a$. 
We will see that $\delta_z =\delta_z(F)=\sqrt{I_p^2-4 Z_{eff} F}$ is a key  
quantity, it controls the tunneling process, and determines the time 
''delay''  under the barrier $\tau_{_{T,d}}$ and the total or the symmetrical 
T-time $\tau_{_{T,sym}}$,  subsec. \ref{ssec:tunt}. 
From fig \ref{fig:ptc} we see that the barrier height at $x_m$:
\begin{equation}\label{hbm}
h_B(x_m)=\mid\!-I_p-V_{eff}(x_m)\!\mid=
\mid\!-I_p+\sqrt{4 Z_{eff}F}\!\mid
\end{equation}
This is the maximum of the barrier height, and by setting $h_B(x_m)=0$ one 
obtains $F_a=I_p^2/(4 Z_{eff})$, which is equivalent to the setting 
$V_{eff}(x_m)=-Ip$ as done by Augst et el \cite{Augst:1989}, and can be easily 
verified from eqs (\ref{hbx}), (\ref{hbm}).
Here we indicate one of the failure of the Keldysh time primary resulting form 
its inadequate definition.
If we recall the definition of the Keldysh time, the time it takes a classical 
electron (with an average velocity) to cross a static barrier of a length 
$l$ \cite{Orlando:2014}. 
For $h_B=0 \Rightarrow l=0$ we get the T-time $\tau_k=0$ (meaning the ejection 
of an electron happen instantaneously at $F=F_a$) because the barrier width 
vanishes for $F\rightarrow F_a$ and 
$\mid\!\!\!x_{e,+}\!\!\!-\!\!\!x_{e,-}\!\!\!\mid=\!0$. 
But we know, at appearance intensity ($I_a= F_a^2$) the ionization time is equal 
to $\frac{1}{I_p}\, (\mbox{in }au)$ \cite{Delone:2000} (chap. 8) and is not zero. 
Which seems naturally because the energy gap that has been overcome is $I_p$,
and as we will see later this follows immediately from our model,  
$\tau_{T,d}(F_a)=\frac{1}{2 I_p}$ and to a total time 
$\tau_{T,sym}(F_a)=\frac{1}{I_p}$, see below. 
As a consequence Keldysh time represents a laboratory clock (parametric time),  
whereas in our following model and time-relation(s) the time is dynamically 
connected to the system (to observe the time form within the system and consider 
the quantum nature of the particle), thus it represents a quantum clock. 
\vspace*{-.50cm}
\subsection{Sketch of the model}\label{ssec:skm}
The (tunneling-) ionization happens according to SFA with 
 zero kinetic energy at the exit point $x_e$. 
Our idea is that the uncertainty in the energy can be quantitatively discerned 
from the atomic potential energy at the exit point 
$\Delta E \sim \mid\! V(x_e)\!\mid=\mid\!-\frac{Z_{eff}}{x_e}\!\mid$ 
for arbitrary field strength $F\le F_a$.  
Then when the electron moves under the barrier in the $x$-direction
 \cite{Klaiber:2013,Yakaboylu:2013}, its kinetic energy getting smaller, at the 
same time it moves upwards on the potential energy scale losing potential energy 
($\frac{-Z_{eff}}{x}$ getting smaller in absolute value). 
The change happens simultaneously in the potential and the kinetic energy while 
staying at the level $-Ip$-line when tunneling, compare fig \ref{fig:ptc}, 
until its kinetic energy becomes zero at the exit point, although  its 
(atomic) potential energy not $\frac{-Z_{eff}}{x_e}\ne 0$.  
This can be also gathered from the analysis of  the short-range Yukawa and 
long-range Coulomb potentials given by Torlina et al. \cite{Torlina:2015}. 
Their conclusion supported with {\it ab initio} numerical tests, 
is that for long range potentials, ionization is not yet completed at 
the ''moment`` the electron exits the tunneling barrier in contrast to 
the usual assumption that ionization is completed once the electron 
emerges from the barrier.
Indeed it is not difficult to see that the electric field of the laser pulse shift 
the electrons along the x-axis direction \cite{Klaiber:2013,Yakaboylu:2013}, 
compare fig \ref{fig:ptc}, reaching the exit point with zero velocity, i.e. the 
electron is forced by the electric field to take and move along a preferred 
direction, and to reduce its kinetic energy to zero at th exit point $x_e$ 
(the field interacts only kinematic-ally with the electron since no photon 
absorbing), where it is still underling the attraction of the atomic potential 
$V(x_e)=-\frac{Z_{eff}}{x_e}$, that defines the uncertainty in the energy and
acts as a shutter open/closed like in the photon box Gedanken experiment with 
an uncertainty proportional to $\Delta E\sim \mid\!V(x_e)\!\mid$. 
As we will see in \ref{sec:dis} the result is very convincing. \\
First it is straightforward to show from eq (\ref{hbx}) that for the classical 
exit point $x_{e,c}$,  setting $\Delta E\sim\mid\!V(x_{e,c})\!\mid$  
leads to incorrect T-time (i.e. it fails to predict T-time measured by the 
experiment). This is because the atomic potential is neglected in calculating 
$x_{e,c}$ \cite{Ivanov:2005}:
\[
 v_e^2-v_0^2 =0-2 I_p=-2 F (x_e-x_0)\Rightarrow x_{e,c} F=I_{p}
\] 
  where $x_e\approx (x_e -x_0)$,  $x_0\approx 0$  is  
the initial point of the electron and assuming the electron moves along  
the x-axis direction \cite{Klaiber:2013,Yakaboylu:2013}.   
It is easy to see, from fig \ref{fig:ptc}, that the ''classical exit`` point 
is the intersection of the electric field line $-x\,F$ with the ionization 
potential of the electron $-Ip$-line, hence $x_{e,c}=\frac{I_p}{F}$, 
which  is far from the ``correct``  exit point. Therefor to use the 
''classical exit`` point, $x_{e,c}$, to determine  the T-time will never give 
a correct answer. It is important to use a correct exit point.  
As  seen in fig \ref{fig:ptc} the relation to other intersection points is simple. 
The crossing points $x_{e,\pm}$ of $V_{_{eff}}(x)$ with the $-I_p$-line are given 
by $h_B(x)=0$, which leads to
\begin{eqnarray}\label{xepm}
x_{e,\pm}=\frac{I_p\pm\delta_z}{2 F}\Rightarrow x_{e,+}=x_{e,c}-x_{e,-}
\end{eqnarray} 
where $\delta_z=\delta_z(F)$ is given in eq (\ref{FFa}), 
we emphasize the dependence of $\delta_z$ on $Z_{eff}$ as done by 
\cite{Dreissigacker:2013}. Note the origin of the axes is at $0$.
We will see later that using the ''classical exit'' point $x_{e,c}$ in the 
uncertainty relation gives the first term in the expansion of the T-time 
obtained using the ''correct'' points $x_{e,\pm}$, eq (\ref{xepm}).  
 
A first arising  question in our model is, what happens in the limit of the 
appearance intensity, i.e. for $F\rightarrow F_a=I_p^2/(4 Z_{eff})$, where  
the electron is shifted from the ground state $E_0=-I_p$ to $E_f\approx0$ 
appearing at $x_{e,-}=x_a=x_{e,+}$ with zero velocity.  
Its  energy uncertainty (since no photon absorption) is then (according to our 
model) $\Delta E(F_a)\sim\mid\!\!-Z_{eff}/x_{a}\!\!\mid=\frac{I_p}{2}$.
 One sees that for atomic field strength ($F_a$) the electron is heavily 
disturbed  but appearing not far from the nucleus at  
$x_a=x_{m}(F_a)=\sqrt{\frac{Z_{_{_{eff}}}}{F_a}}=\frac{2 Z_{eff}}{I_{p}}$ 
with an ionization time that follows immediately from TEUR in eq (\ref{ucr}): 
\begin{equation}\label{Tta}
 \tau_a=\frac{1}{2\mid\!\!\Delta E(F_a)\!\!\mid}=\frac{1}{I_p}
\end{equation}
as it should be for the ionization process at the $F_a$ \cite{Delone:2000} 
(chap. 8). However we will see later that $F_a$ is a special case because 
 $x_{e,-}=x_a=x_{e,+}$ (a double solution of $h_B(x)=0$) and  the limit 
$\tau_{T,sym}(F\rightarrow F_a)=\frac{1}{I_p}$ is a sum of two equally terms 
$\frac{1}{2I_p}$ (see sec. \ref{ssec:tunt}, and  \ref {sec:pr1}). 
In conclusion, our model is meaningful and the atomic potential energy of 
the electron at the tunnel exit $x_e$ (instead of the gravitational potential 
in {\it the Einstein-Bohr Gedanken experiment)} amounts to calculate the 
uncertainty of the energy in the tunneling process, and hence the T-time
by the virtue of eq (\ref{ucr}) which leads to an excellent result as we 
will see sec. \ref{sec:dis}.
\vspace*{-0.50cm}
\subsection{Tunneling time}\label{ssec:tunt} 
Our goal now to find an expression to calculate the T-time, and  what we need is 
the correct exit point, where  many approximations exist. The most used 
one in the literature can be obtained from eq (\ref{hbx}), where the barrier 
height at the entrance and the exit vanishes $h_B(x_{e,\pm})=0$, they are the 
crossing points of $V_{eff}$ curves with the $-I_p$-line, see fig \ref{fig:ptc}, 
\begin{eqnarray}\label{hbx0}
h_{B}(x)=0\Leftrightarrow -\frac{Z_{eff}}{x}-x F=-I_p
\end{eqnarray}
Solving eq (\ref{hbx0}) gives immediately eq (\ref{xepm}).   
As seen in fig \ref{fig:ptc}, $x_{e,-}$ is the inner crossing point, the 
``entrance'' point, and  $x_{e,+}$ is the outer crossing point, 
the  ``exit`` point.  
Physically is argued that the electron escapes the barrier at $x_{e,+}$, when 
it moves in the direction $x_{e,-}\rightarrow x_{e,+}$ and vice versus for the 
opposite direction, and we will see this presents a useful symmetry property 
of the tunneling process when deriving the T-time.
 
Now we can calculate the uncertainty in the energy $\Delta E(x_{e,+})$ by using 
the exit point $x_{e,+}$ (in the direction $x_{e,-}\rightarrow x_{e,+}$) and 
from this the T-time. 
From eqs (\ref{xepm}), (\ref{ucr}) and according to our model we get:
\begin{eqnarray}\label{dE1}
\Delta E(x_{e,+})&=&\mid\!\!\frac{-Z_{eff}}{x_{e,+}}\!\!\mid
=\frac{Z_{eff}\cdot 2F}{(I_p+\delta_z)}
= \frac{(I_p-\delta_z)}{2} \quad\\ 
\tau_{_T,unsy} &=&\frac{1}{2}\frac{1}{\Delta E}=\frac{1}{(I_p-\delta_z)}\label{Tunsy}
\end{eqnarray}
which we call the unsymmetrical T-time $\tau_{T,unsy}$. We will see later that 
a factor (1/2) is missing that can be recovered by a symmetry consideration.   
We first show that the first order of eq (\ref{Tunsy}) is equal to the T-time 
resulting from using $x_{e,c}$, the ''classical exit'' point, and then we look 
to the symmetry of the tunneling process.    
Expanding eq (\ref{Tunsy}) in term of $\xi=(\frac{4 F}{I_p^2})$, we get in 
the first order  $\tau_{_{T,c}}$, the T-time at the ``classical exit'' point 
$x_{e,c}$ as mentioned above, then 
\begin{eqnarray}\label{Tc}
 O^1(\tau_{_T,unsy})=\frac{I_p}{2 F}=
\frac{1}{2}\frac{1}{\Delta E{_{c}}}= \tau_{_{T,c}}
\end{eqnarray}
where $\Delta E_{c}=\mid\!\!\frac{-1}{x_{e,c}}\!\!\mid$.
What about the inner point $x_{e,-}$? we could
 assume, due to the $\delta_z$-symmetry between $x_{e,+},\, x_{e,-}$, that 
the electron enter the barrier backwards from $x_{e,+}$ ''entrance``  
to $x_{e,-}$ ''exit'' with an uncertainty (according to our model) 
$\Delta E(x_{e,-})=\mid\!\!\frac{-Z_{eff}}{x_{e,-}}\!\!\mid$, which leads to
(compare eq (\ref{dE1}), (\ref{Tunsy})) 
\begin{equation}\label{Tunsy1}
\frac{1}{2\Delta E(x_{e,-})}=
\frac{(I_p-\delta_z)}{4 Z_{eff}\cdot F}
= \frac{1}{(I_p+\delta_z)}
\end{equation}
This symmetry is deduced in a similar way the Aharanov-Bohm time operator 
\cite{Aharonov:1961} is defined for  a free particle  
$\hat T= \frac{1}{2} (\hat{x} \, \hat{p}^{-1} + \hat{p}^{-1} \hat{x})$ 
or in more elaborate (the so-called bilinear form) and detailed treatment given 
by Olkhovsky and Recami \cite{Olkhovsky:2009,Olkhovsky:1970}. 
Such  operators (given by Aharanov-Bohm or Olkhovsky) have the property of 
maximally symmetric in the case of the continuous energy spectra, and the 
property of quasi-self-adjoint 
\footnote{it can be chosen as an almost self-adjoint operator with practically 
almost any degree of the accuracy \cite{Olkhovsky:2009}} 
operators in the case of the discrete energy 
spectra \cite{Olkhovsky:2009} (and the references therein), and are the nearest 
best thing to  self-adjoined operators and satisfy the conjugate relation with 
the Hamiltonian and therefore implies an ordinary TEUR \cite{Olkhovsky:2009}  
and \cite{Muga:2008} (chap. 1). 
We use this property, i.e. we assume that the maximally symmetric 
(or almost self-adjoint \cite{Olkhovsky:1970}) property holds for the T-time, 
for more details see \cite{Olkhovsky:2009}, which leads using eqs (\ref{Tunsy}), 
(\ref{Tunsy1}) to a simple relation for what 
we call the symmetrical T-time given by: 
\begin{eqnarray}\label{Tsym}
\hspace*{-0.5cm}\tau_{_T,sym}&=&\tau_{_{T,+}}+\tau_{_{T,-}}
=\tau_{_{T,i}}+\tau_{_{T,d}}\\\nonumber
&=&\frac{1}{2}\left(\frac{1}{\Delta E^-}+\frac{1}{\Delta E^+}\right)\\\nonumber
&=&\frac{1}{2}\left(\frac{1}{(Ip+\delta_z)}
+\frac{1}{(Ip-\delta_z)}\right)=\frac{I_p}{4Z_{eff}F}
\end{eqnarray}
where we defined $\tau_{_{T,\pm}}=1/(2\Delta E^{\mp})={(2(Ip\pm\delta_z))}^{-1}$, 
or $(1/2)\Delta E^{\pm}= \Delta E(x_{e,{\pm}})$. 
We call the first term of eq (\ref{Tsym}) $\tau_{_{T,i}}(=\tau_{_{T,+}})$ and the 
second term $\tau_{_{T,d}}(=\tau_{_{T,-}})$ for a reason that will be clear in 
the next section, sec. \ref{sec:pr}. 
Relation (\ref{Tsym}) has again (clearly because $\delta_z=0)$ the correct limit 
for atomic field strength (compare eq (\ref{Tta}) and the discussion after it): 
\[
\tau_{_T,sys}(F\rightarrow F_a)=\frac{1}{2Ip}+\frac{1}{2Ip}
=\frac{I_p}{4\,Z_{eff}\frac{Ip^2}{4\,Z_{eff}}}=
 \frac{1}{Ip}
\] 
Note that the limit $F\rightarrow F_a$ gives $x_{e,+}=x_{e,-}=x_m(F_a)=x_a$ which 
means that the two points coincide at the top of the barrier (a double solution 
of eq \ref{hbx0}). 
The question is whether this means a symmetry break of the tunneling process? 
because at this limit there is only one direction, namely towards the continuum, 
so that  the ``tunneling`` becomes a ''real''  ionization 
(or an ejection) process at the appearance intensity $I_a=F_a^2$, and  $\delta_z$ 
becomes imaginary for super-atomic field strength $F>F_a$ (whereas $F<F_a$ is 
called the subatomic field strength), see below sec \ref{sec:dis}. 
A reasonable question in this case, whether the time to reach the ``entrance`` 
of the barrier and under the barrier, intrinsic or dynamical to be measured 
by a quantum clock, becomes after the tunneling a classical, external 
(or parametric) time due to the break of some symmetry property? so that 
(only) under such a symmetry break a quantum clock (the internal time) coincides 
with a laboratory clock (the external time). 
     
In sec. \ref{sec:dis} we will see that eq (\ref{Tsym}) (especially $\tau_{T,d})$ 
gives an excellent agreement with the experimental result of \cite{Eckle:2008s}.   
It worthwhile to mention that the T-time measured by the experiment 
is the  time that the electron spends when it moves under 
the barrier from $x_{e,-}$ to $x_{e,+}$, which corresponds to the  second term 
of eq (\ref{Tsym}) $\tau_{_{T,d}}=\frac{1}{2}\frac{1}{(Ip-\delta_z)}$, 
see further sec. \ref{sec:dis} result and discussion. 
\vspace*{-0.50cm}
\section{Physical reasoning}\label{sec:pr}
\vspace*{-0.25cm}
\subsubsection{Tunneling time and a model of a shutter}\label{sec:pr1}
\vspace{-0.25cm} 
The theoretical mathematical model developed in sec. \ref{sec:TM} 
can be supported and derived by physical arguments, the only difference is 
 that we try to figure a  physical insight that helps us to  put physics in 
mathematical relations.
From fig \ref{fig:ptc} and eq (\ref{xepm}) we get ($d_B(F)$ is the barrier width):
\begin{equation}\label{delF}
d_B(F)=x_{e,+} - x_{e,-}=\frac{(I_p+\delta_z)}{2F}-\frac{(I_p-\delta_z)}{2F}
=\frac{\delta_z}{F}
\end{equation}
from this it follows  $\delta_z=(x_{e,+} - x_{e,-})\, F$. Because for atomic 
field strength $F=F_a\Rightarrow \delta_z=0,\, d_B=0$, we can interpret   
$\delta_z$ as  the field's (kinetic) energy exerting in the tunneling process 
between the ``entrance`` (or the ``inner``) point and the ''outer`` exit point.
The uncertainty in the energy as the electron moves on the $-Ip$-level to 
$x_{e,+}$, i.e. when tunneling the barrier and escapes at the exit to the 
continuum is then $\Delta E^+=abs(-I_p+\delta_z)$.  
That means the barrier itself causes a delaying time relative to the atomic field 
strength $F_a$ ($\delta_z =0$). The T-time is then obtained from eq (\ref{ucr}), 
\begin{equation}\label{Td}
\hspace{-0.25cm}\tau_{_{T,d}}\equiv\frac{1}{2 \Delta E^+}
=\frac{1}{2\,(I_p-\delta_z)}\,(\mbox{for } F\le F_a)
\end{equation}
which we call $\tau_{_{T,d}}$ (compare eq \ref{Tsym}) meaning that is the time 
duration (time interval) to cross the distance under the barrier 
(in the direction $x_{{_e,-}}\!\rightarrow\! x_{_{e,+}}$) and escapes at 
the exit point $x_{e,+}$ to the continuum. 
The term $\tau_{T,d}(F\rightarrow F_a)=\frac{1}{2I_p}$ at the limit $F=F_a$ 
account for the turning off the wave packet (or shack-off step, see discussion 
below) at the ''entrance-exit'' point $x_a$ to escape to the continuum, which 
indicates that the shack-off step or turning off at $x_{e,+}$ for 
$F<F_a$ happens with a time delay as given in eq (\ref{Td}). 
Expanding $\delta_z=Ip\sqrt{1-(4Z_{eff} F/Ip^2)}$ and taking the first 
we get eq (\ref{Tsym}), $\tau\approx\frac{I_p}{4 Zeff F}=\tau_{_{T,sym}}$. 
It means using the symmetrization gives a linearized time duration 
(linearized T-time).  
As we already mentioned the limit of eq (\ref{Td}), $1/(2Ip)$, 
 is only one part of the total tunneling time $1/Ip$. 

Now we argue that, this picture 
fits well in the Gedanken experiment of Einstein and the (intrinsic) time 
$\tau_{_{T,d}}$ (the second term in eq (\ref{Tsym}) or (\ref{Tt})) is rather 
the time of passage with the shutter open/closed time interval (generated in the 
internal time \cite{Aharonov:2000}) is related by the virtue of eq (\ref{ucr}) 
to the uncertainty $(1/2)\Delta E^+=\Delta E=V(x_{e,+})$ (note we recovered in 
eq (\ref{Tsym}), (\ref{Td}) the factor (1/2) missing in eq (\ref{Tunsy})), 
and we think that the attosecond experiment with the help of our model represents 
a realization of the {\it photon-box Gedanken experiment} (with the electron as 
a particle instead of the photon) with an uncertainty being determined from the 
(Coulomb) atomic potential due to the disturbing by the field $F$, instead of 
(the disturbing by) the wighting process and as a result an uncertainty in 
the gravitational potential \cite{Aharonov:2000}, as done by the famous prove 
of Bohr to the uncertainty (or indeterminacy) of time in the 
{\it photon-box Gedanken experiment} 
\cite{Aharonov:2000,Busch:1990I,Busch:1990II}.
\vspace{-0.5cm}
\subsubsection{Totale time. ($F=F_a$)} 
\vspace{-0.25cm}
At the moment one can obtain the total time, i.e. including the time to reach 
the entrance of the barrier $x_{e,-}$, by adding the term $1/(2I_p)$
\begin{eqnarray}\label{Tt}
\tau_{_{T,t}}&\approx&
\frac{1}{2}\left(\frac{1}{I_p}+\frac{1}{abs(-I_p+\delta_z)}\right)\\\nonumber 
 &=&\frac{1}{2}\left(\frac{1}{I_p}+\frac{1}{I_p-\delta_z}\right)
\end{eqnarray}
where the index $t$ only to distinguish between different notations.
Nevertheless the term $1/(2I_p)$ we added is exact only when $F=F_a$ or
$x_{e,-}(F_a)=x_a=x_{e,+}(F_a)$,
hence we call it $\tau(x_i,x_a)$ 
\begin{equation}\label{Txi}
\tau(x_i,x_a)=\frac{1}{2I_p}
\end{equation} 
where  $x_i$ is the initial point. It can be viewed as the response time of the 
electron to the field, that is, the electron received a kick by the field, and is 
polarized along the field direction, and (jumping up \cite{Ivanov:2005}) is moving 
from $x_i$ to the ``entrance-exit``  point $x_a$ to the continuum, 
$x_i\rightarrow x_{e,-}(=x_a=x_{e,+}) \rightarrow\infty$. 
In this case ($h_B(F=F_a)=0$, $\delta_z=0$ ) the most probable ''tunneling'' 
mechanism is that the laser field distorts the electron, shakes it up (moving from 
$x_i$ to $x_a$), and shakes it off (moving to the continuum) at a (total) time 
given in eqs (\ref{Tsym}) or (\ref{Tt}) 
$\tau=\frac{1}{2}\left(\frac{1}{I_p}+\frac{1}{I_p}\right)=\frac{1}{I_p}$.
In this model, for $F=F_a$ the (illustrative) two steps are not strictly 
separated, whereas, see further below, for $F<F_a$ they are well separated 
due to the barrier $d_B(F)>0=d_B(F_a)$. 
\vspace{-0.50cm} 
\subsubsection{Total time for subatomic field $F< F_a$}
\vspace{-0.25cm} 
However, $x_a=x_{e,-}=x_{e,+}$ is the maximum ``entrance`` point (most right see 
fig \ref{fig:ptc}), the electron is less disturbed for  $F<F_a$ and moved to a 
point $x_{e,-}<x_a$  that is  closer to the initial point $x_i$
\footnote{Note that for small field strength $F<<F_a$ the electron absorbs one or 
more photons and moves vertically on the energy scale emerging at exit point very 
close to the initial point $x_i$ on the coordinate scale}, this shortens the time 
to reach the ``entrance`` point, and we expect that the response 
of the electron to a small field strength $F$ is weaker than to a stronger field 
$F\rightarrow F_a$. 
The time reduction in $\tau(x_i,x_{e,-})$ for $F$ comparing 
to $\tau(x_i,x_a)=1/(2I_p)$ for $F_a$, eq (\ref{Txi}), is a factor depending 
on $\delta_z$ (see discussion after eq (\ref{delF})), because the kinetic energy 
experiences a change proportional to $(x_{_{e,-}}-x_i)\, F< (x_a-x_i) \, F_a$. 
A (weaker) field $F<F_a$ is not sufficient to compensate for the kinetic energy 
at the (shake-up) step,  
instead the electron is at the ''entrance'' $x_{e,-}<x_a$ with a velocity that 
is sufficient to inter under the barrier and reaches the exit point $x_{e,+}$ 
with zero velocity. 
Here is again an another failure of the Keldysh time (see also discussion after 
eq (\ref{hbm})) that is the electron did not inter the barrier with its initial 
velocity $\sqrt{2 I_p}$ suggested by many authors \cite{Ivanov:2005}. 
Now we give the following relation for the time needed to  reach the 
''entrance`` point $x_{e,-}$, and show an explanation further below.
\begin{equation}\label{Ti}
\tau_{T,i}\equiv \tau_{(x_i,x_{e,-})}=\frac{1}{2(I_p+\delta_z)}\equiv
\frac{1}{2\Delta E^-} 
\end{equation}
The factor $\delta_z$ (comparing to eq (\ref{Txi}) for $F_a$) in the denominator 
results from two parts.
Indeed we follow \cite{Aharonov:2000} in that the uncertainty is a result of 
the different reactions or responses of the electron to different filed strengths.
The one part comes from the difference of moving along the $x-$axis. i.e. 
the difference in shifting the electron to 
$x_a$ with $F_a$ or to $x_{e,-}<x_a$ for  $F<F_a$, which leads to 
$\Delta_1=x_a F_a -x_{e,-} F =\frac{I_P}{2}-\frac{(I_p-\delta)}{2}=
\frac{\delta_z}{2}$. 
The other part can be deduced from the change on the vertical 
(potential energy) scale.
 When the electron receives a kick, changing its potential (on a vertical scale), 
its atomic potential has a different changes  between $V(x_i)$ and $V(x)$ for 
$x=x_a$ or $x_{e,-}$. But this is a result of different responses (on energy 
scale), while the electron is forced to follow an orientation along the 
(opposite) field direction at $x_a, \mbox{ or at } x_{e,-}$. 
Therefor this part can be approximated from the difference $\Delta V(x)$ in 
the atomic potential at the different ``entrance` points,  which gives 
$\Delta_2=V(x_a)- V(x_{e,-})
=-\sqrt{Z_{eff}F_a}+\frac{2 Z_{eff} F}{(I_p-\delta_z)}=
-\frac{I_p}{2}+\frac{(I_p+\delta_z)}{2}=\frac{\delta_z}{2}$.
We are led to a difference equal to $\Delta_1 + \Delta_2=\delta_z$ between 
reaching the ''entrance'' point $x_{e,-}$ (for $F$) relative to  the 
``entrance-exit`` point $x_a$ (for $F_a$), leading to an energy uncertainty  
$\Delta E=abs(-I_p-\delta_z)=(I_b+\delta),\mbox{ for}\, F\le F_a$, 
and  with a time  
$\tau_{T,i}(F)=\frac{1}{2(I_p+\delta_z)}$ for arbitrary subatomic field 
strength $F$, 
hence one obtains eq (\ref{Ti}) instead of eq. (\ref{Txi}).    
We have explained so far eq (\ref{Td}) and eq (\ref{Ti}), in doing so we 
explained the physical meaning of the symmetry consideration done above 
(see eq (\ref{Tsym}) and the discussion before) and we obtain from 
eqs (\ref{Td}), (\ref{Ti}) the result obtained in eq (\ref{Tsym}):
\begin{eqnarray}\label{Tsym1}
\tau_{_{T,sym}}&=& \tau_{_{T,i}}+\tau_{_{T,d}}\equiv\tau_{_{T,+}}+\tau_{_{T,-}}
\\\nonumber
&=&
\frac{1}{2}\left(\frac{1}{I_p+\delta_z}+\frac{1}{(I_p-\delta_z)}\right)
=\frac{I_p}{4Z_{eff} F}
\end{eqnarray} 
the first term $\tau_{_{T,+}}=1/(2E^-)$ corresponds to the first step, where 
the electron is shacked up and moved to the ''entrance`` $x_{e,-}$ (or $x_a$ 
for $F_a$) that takes the times $\tau_{_{T,i}}=\tau_{_{T,+}}$.  
The second term  $\tau_{_{T,-}}=1/(2E^+)$ corresponds to the actual  
T-time or moving under the barrier and shaken off to the continuum, 
with a time delaying $\tau_{_{T,d}}=\tau_{_{T,-}}$ 
due to the barrier relative to the atomic field strength $F_a$, where $\delta_z=0$ 
and $\tau_{_{d,T}}=\frac{1}{2I_p}$.  

For $F_a$, as already mentioned, the second or shake-off step  
follows immediately the first or shack-up step  
and the two steps are not strictly separated.
For $F<F_a$ the two steps model of the tunneling process are well separated.
They happen with opposite directions at the time scale, the first 
step is less time consuming since $F$ causes a smaller disturbance relative to 
$F_a$, and the the electron moves not far from its initial position for small F,  
$x_{e,-}<x_a$, whereas the second step happens with a time 
delaying, $x_{e,+}> x_a$, relative to the ionization at atomic field strength.  
So far our theoretical model is assisted with an explanation through a physical 
reasoning. In the following we show and discuss our result for He-atom with
 a comparison to the experiment \cite{Eckle:2008,Eckle:2008s}.
\vspace*{-0.5cm}
\section{Result and discussion}\label{sec:dis}
\begin{figure}[t] 
\vspace{-4.50cm}
\hspace{-2.0cm}\includegraphics[height=15.0cm,width=11cm]{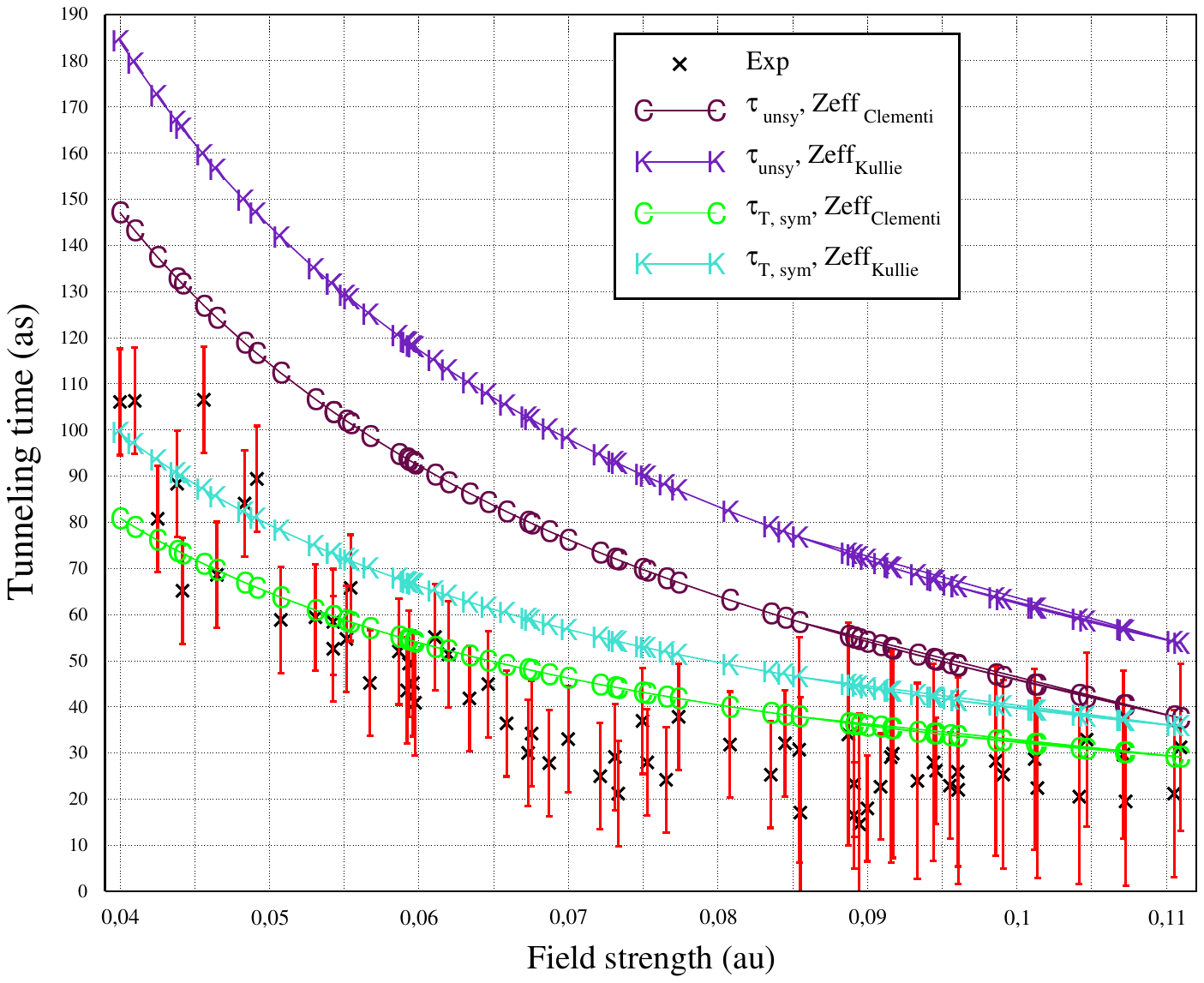}
\vspace{-5.0cm}
\caption{\label{fig:tut}\tiny
T-time  $\tau_{_{T,unsy}}$ eq (\ref{Tunsy}), and  $\tau_{_{T,sym}}$ eq (\ref{Tsym}),
for two $Z_{eff}$ models \cite{Kullie:1997} and \cite{Clementi:1963}. 
Time is in attosecond units vs laser field strength in atomic 
units, corresponds to the tunneling ionization of the He-Atom in strong field, 
in good agreement with experimental result 
\cite{Landsman:2014II,Eckle:2008s,Eckle:2008}. 
Experimental values are kindly sent by A. Landsman \cite{Landsman:2014II}.}
\end{figure}

\begin{figure}[t]  
\vspace{-4.50cm}
\includegraphics[height=15.0cm,width=11cm]{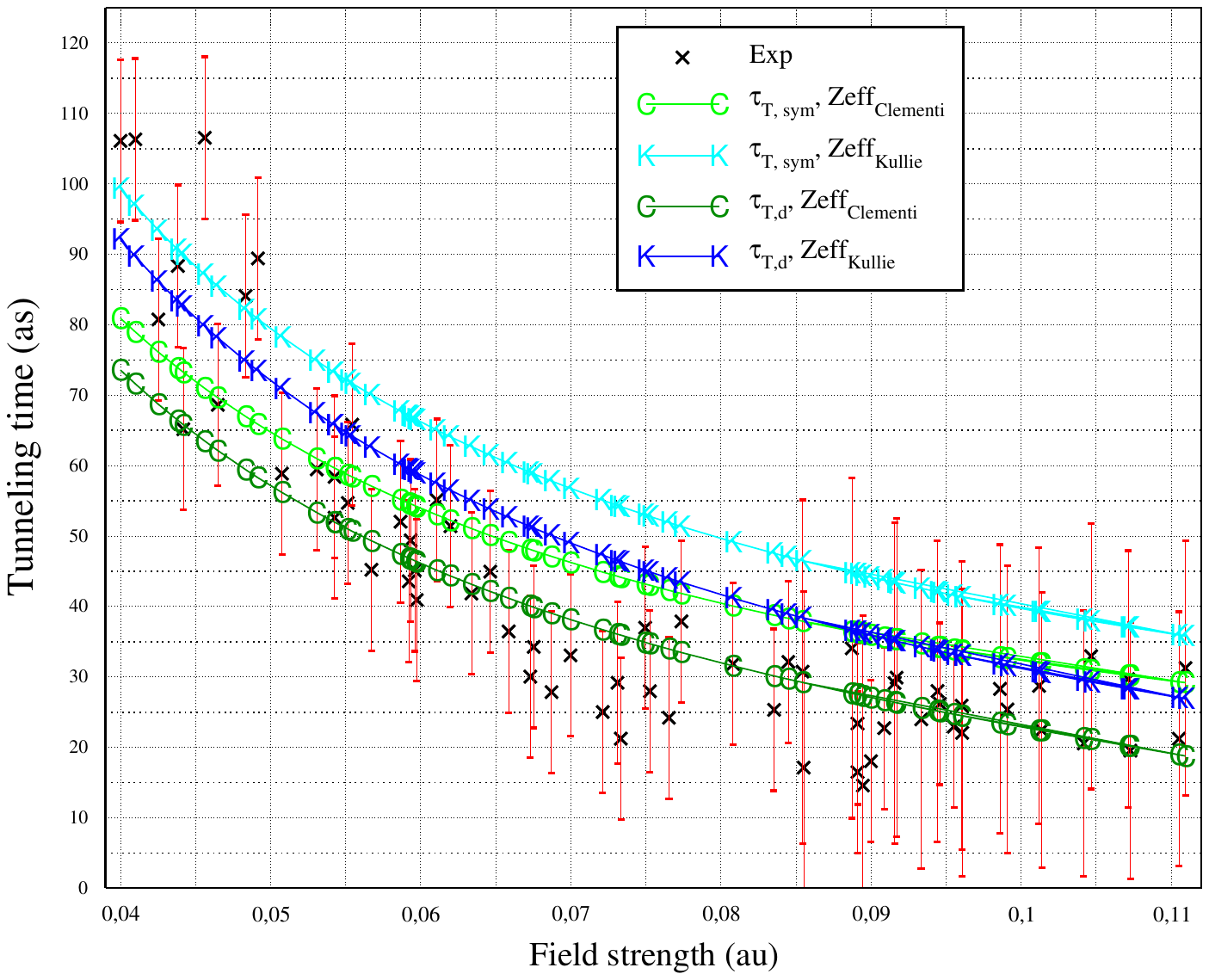}
\vspace{-5.0cm}
\caption{\label{fig:tut1}\tiny 
T-time  $\tau_{_{T,d}}$ eq (\ref{Td}) and $\tau_{_{T,sym}}$ eq (\ref{Tsym1}), 
(note eq (\ref{Tsym1}) is identical to eq (\ref{Tsym}), see fig \ref{fig:tut}),  
for two different $Z_{eff}$ models as in fig \ref{fig:tut}.  
Time vs laser field strength as in fig \ref{fig:tut}, corresponds to the tunneling 
ionization of the He-Atom in strong field, in excellent agreement with experimental 
result \cite{Landsman:2014II,Eckle:2008s,Eckle:2008}. 
Experimental values as in fig \ref{fig:tut}.}
\end{figure}

\vspace*{-0.25cm}
In fig \ref{fig:tut} we show the results of eq (\ref{Tunsy}) the unsymmetrical 
$\tau_{_{T,unsy}}$, and eq (\ref{Tsym}) the symmetrical  T-time $\tau_{_{T,sym}}$. 
The results for $\tau_{_{T,d}}$ eq (\ref{Td}), and again the the symmetrical 
(or total) T-time $\tau_{_{T,sym}}$ of eq (\ref{Tsym1}) are shown in 
fig \ref{fig:tut1}. 
Note eq (\ref{Tsym1}) is identical with eq (\ref{Tsym}), whereas eq (\ref{Td}), 
is the second term of eq (\ref{Tsym1}) or (\ref{Tsym}), which is the actual T-time, 
i.e. the time needed to pass the under barrier region $(x_{e,-}\rightarrow x_{e,+})$ 
and escape at the exit point to the continuum, and that is usually the T-time 
measured in the experiment. 
The results are for the He-atom in a comparison with the experimental result of  
\cite{Eckle:2008s}. 
The experimental data and the error bars in the figure were kindly sent by 
A. Landsman \cite{Landsman:2014II}.   
We plotted the relations (\ref{Tunsy}), (\ref{Tsym}), (\ref{Td}) and (\ref{Tsym1}) 
at the values of the field strength at maximum of the elliptically polarized 
laser pulse ($\lambda =735$, elliptical parameter  
$\varepsilon=0.87$, $F=F_0/\sqrt{1+\varepsilon^2}$) used by the experiment 
exactly as given in \cite{Landsman:2014II}.   
In fig \ref{fig:tut}, the upper two curves are the unsymmetrical T-time 
$\tau_{_{T,unsy}}$, eq (\ref{Tunsy}), for two different models of the effective 
nuclear charge $Z_{eff}$, that the tunneling electron experiences during the 
tunneling process. 
Accordingly the lower two curves are the symmetrical T-time, $\tau_{_{T,sym}}$,
 eq (\ref{Tsym}) for the two different models of $Z_{eff}$. 
We mention that eq (\ref{Tt}) (not plotted) gives a closer result to eqs 
(\ref{Tsym}), (\ref{Tsym1}). 
The two different effective charge models are from Kullie \cite{Kullie:1997},  
with $Z_{_{eff,K}}=1.375$  and Clementi \cite{Clementi:1963} with 
$Z_{_{eff,C}}=1.6875$ . 
In fig \ref{fig:tut} we see that our $\tau_{_{T,unsy}}$ is not close to 
the experimental data of \cite{Eckle:2008s}. Whereas $\tau_{_{T,sym}}$ is close 
for both models of the $Z_{eff}$. But we notice, see discussion further below, 
that the first term in eqs (\ref{Tsym}), (\ref{Tsym1}) is much smaller than the 
second $\tau_{_{T,i}}<\tau_{_{T,d}}$ for small $F$ (relative to $F_a$).  
 
Concerning $Z_{eff}$  we see for small field strength $F\lesssim 0.05$ 
that $\tau_{_{T,sym}}$ with $Z_{_{eff,K}}$ is closer to the experimental data 
(and especially for $\tau_{_{T,d}}$, see discussion below fig \ref{fig:tut1}). 
The reason is that the $Z_{_{eff,K}}$ model is a H-atom-like model, based on the 
assumption that the first electron of the He-atom occupies the $1s$-orbital 
(with probability density $\mid\!\!\Psi(r)\!\!\mid^2$, which screens the the 
nuclear charge and the second electron is treated as an active electron 
or a ``valence`` electron \cite{Kullie:1997} similar  
to the SAE approximation. 
This is a good approximation when the tunneling electron moves far from 
the left atomic core (He$^+$), or  the barrier width is large 
$x_{e,+}\!>\!15\, au$, hence the better agreement, and possibly 
this is important for smaller field strength in the region where 
$\gamma_{K}\approx 1$. 
In the range of larger field strength  multielectron effects are 
expected and the model $Z_{_{eff,C}}$ based on Hartree-Fock calculation is
 more reliable,  where the electron moves not far from the left atomic core
(small barrier width) and hence the better agreement in this region.
It is likely that a model depending on the $x$-coordinate $Z_{eff}(x_{e,+})$ 
will achieve a better agreement that smoothly fits the two regions.

Now we look to fig \ref{fig:tut1}, where $\tau_{_{T,sym}}$ eq (\ref{Tsym1}) and  
$\tau_{_{T,d}}$ eq (\ref{Td}) are shown.
Eq (\ref{Tsym1}) is the same as eq (\ref{Tsym}) (shown in fig \ref{fig:tut}). 
For the $\tau_{_{T,d}}$ we see an excellent agreement with the 
experiment. As already mentioned $\tau_{_{T,d}}$ corresponds to the T-time 
measured in the experiment, that is the time (interval) needed to move under 
the barrier from  the entrance to the exit point and escape to the continuum 
with a shack-off, or between the instant of orientation at $x_{e,-}$ an the 
instant of ionization at $x_{e,+}$, which is the time spent in the classically 
forbidden region \cite{Eckle:2008s}. 
For small $F\lesssim 0.055\, au$,  $Z_{eff,K}$ 
gives better agreement with the experiment, whereas for larger field strength 
$Z_{eff,C}$ is more reliable, where multielectron effects are expected due to 
the decreasing  width of the barrier and the tunneling electron is closer to the 
first one when it traverses the barrier. 

In fig \ref{fig:tut1} we see that the difference between the total or symmetrical 
T-time $\tau_{_{T,sym}}$ and the (actual) T-time $\tau_{_{T,d}}$ is small, because 
the second term  $\tau_{_{T,d}}$ in eq (\ref{Tsym1}), 
incorporates the delaying time caused by the barrier and is the main time of 
the tunneling process for large barrier. 
Whereas the first part term  $\tau_{_{T,i}}$, is due to the shake-up 
of the electron by the field moving it to the ''entrance'' $x_{e,-}$, 
which is small for small $F$.
For large field strength the two parts become closer because the barrier width 
getting smaller $\delta_z/F=(x_{e,+}-x_{e,-}) \rightarrow 0$ and for appearance 
intensity ($\delta_z =0)$ they become equal.  
Concerning $Z_{eff}$ in fig \ref{fig:tut1}, we readily see for $\tau_{_{T,d}}$ 
the same behavior as in fig \ref{fig:tut}.

\begin{figure}[t]  
\vspace{-5.20cm}
\hspace*{-0.250cm}\includegraphics[height=13.225cm,width=11.cm]{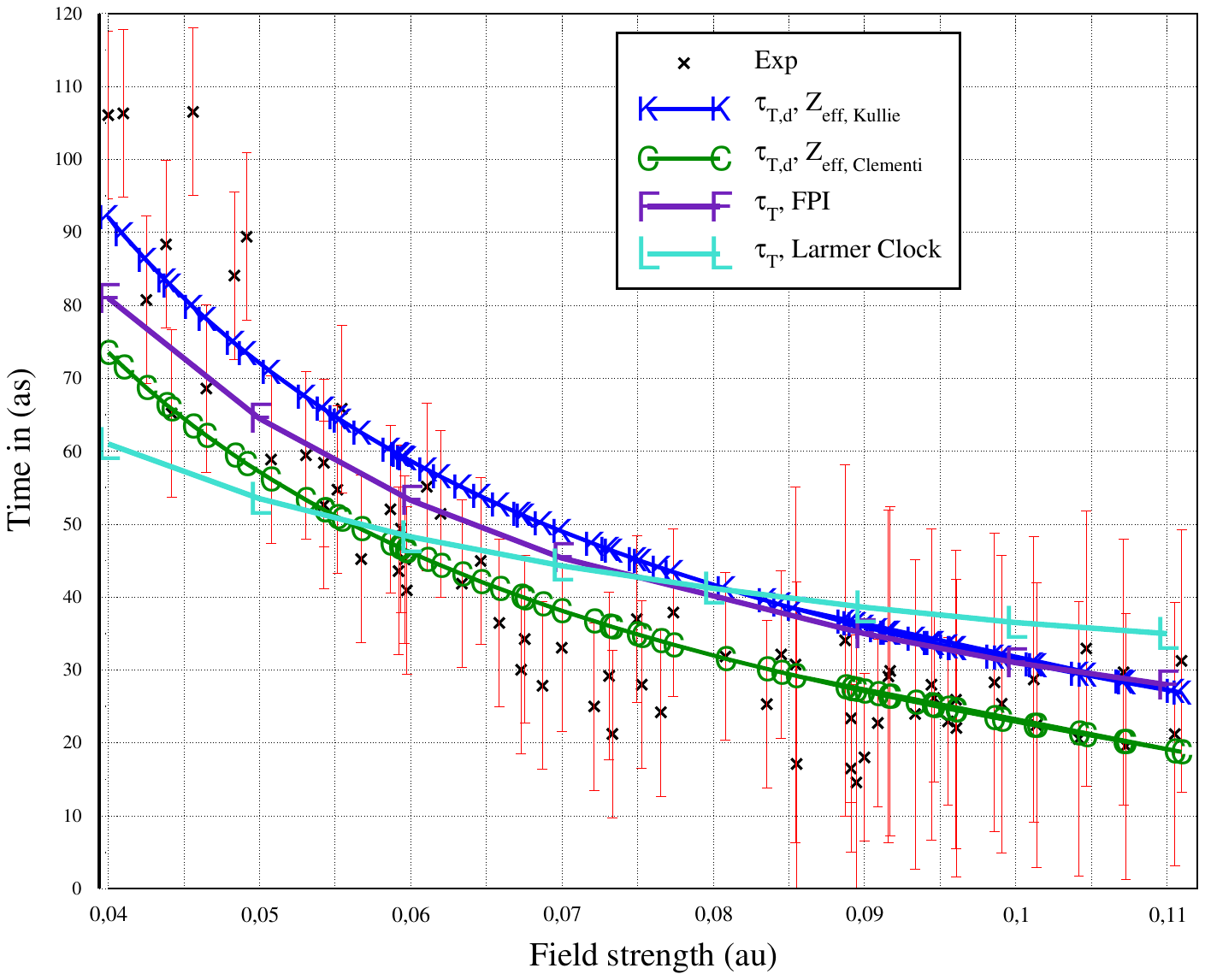}
\vspace{-4.cm}
\caption{\label{fig:FPI}\tiny 
T-time  $\tau_{_{T,d}}$ eq (\ref{Td}), for two different $Z_{eff}$ models as 
in fig \ref{fig:tut} together with FPI result \cite{Landsman:2014II} and the 
experimental result \cite{Eckle:2008s,Eckle:2008}.
Experimental values and  FPI result are kindly sent by A. Landsman and C. Hoffman 
\cite{Landsman:2014II,Hofmann:2013}.} 
\end{figure}

In fig \ref{fig:FPI} we plotted our result $\tau_{_{T,d}}$ eq(\ref{Td}) together
with the FPI result of \cite{Landsman:2014II} (data were kindly sent by A. Landsman 
and C. Hofmann \cite{Landsman:2014II,Hofmann:2013}). 
From the figure we see a good agreement between the two results, and the difference 
is smaller than the experimental error bars.
Indeed we expect that the FPI would agree better for large field strength $F>0.055$ 
with the lower curve ($Z_{eff,C}$, green). For small field strength, FPI is more 
or less close to both curves (green, blue), but our upper curve ($Z_{eff,k}$, blue) 
tends to be in a better agreement with the experimental values for smaller 
field strength.  
An important point is that our model and result(s) 
predict a (real) T-time (time of passage) of a single particle, it is not 
distributive of an ensemble  (although indeterminately in regard to the 
uncertainty relation) and we make no assumption about the path of the particle 
inside the barrier, whereas the FPI treatment is probabilistic/distributive 
that makes a use of all possible (classical) paths inside the barrier that have 
a traversal tunneling time $\tau$. 
Furthermore, Landsman et al \cite{Landsman:2014I} uses the time $\tau_0$, 
which is determined by the measurement to coarse-grained the probability 
distribution of the T-times to achieve the aimed results, although Sokolovski 
\cite{Sokolovski:1990}, \cite{Muga:2008} (chap. 7) claims (in regard to his FPI 
description) that no real time is associated with the tunneling.  
We think that the two views are rather complementary as it is usual in quantum 
mechanics: wave/particle or individual (single particle)/statistical 
(distributive) etc.
The Larmer Clock should in principle agrees with our result, the result of 
Landsman \cite{Landsman:2014II} shows that the agreement is good only for 
$F\approx 0.06-0.1$, hence Larmer Clock values are inferior for  $F<0.06$. 
The same holds for $F>0.1$, although the difference is smaller than the error 
bars, but the trends of of Larmer Clock curve for large field strength seems not 
correct. In general the Larmer Clock curve is flat comparing with the other 
curves and the experimental data, which is difficult to understand.   

\begin{figure}[t]  
\vspace{-6.00cm}
\includegraphics[height=15.0cm,width=11cm]{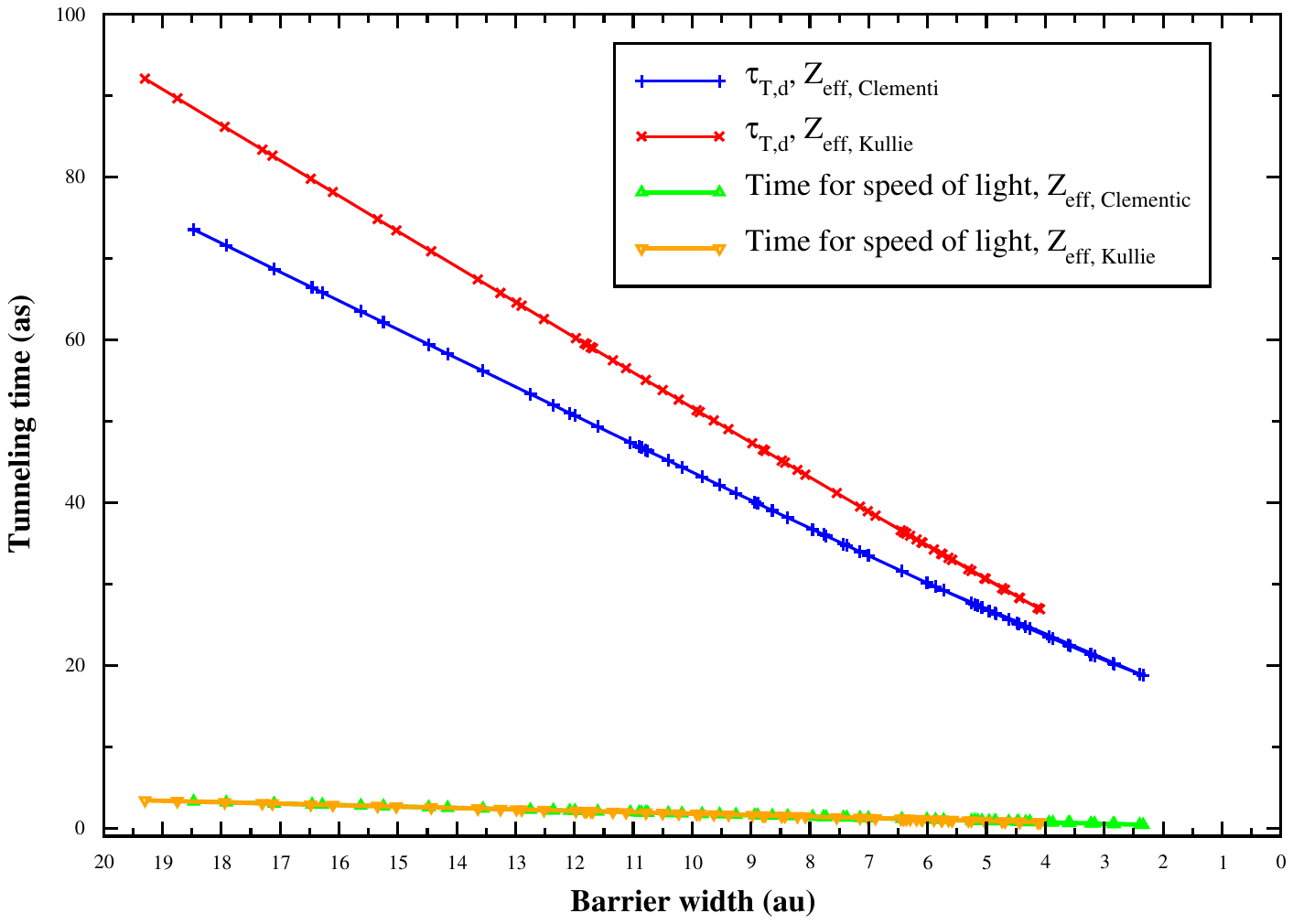}
\vspace{-5.50cm}
\caption{\label{fig:tbwd}\tiny 
T-time  $\tau_{_{T,d}}$ eq (\ref{Td}) in as vs barrier width $d_B(F)$ in au.  
for two different $Z_{eff}$ models as in fig \ref{fig:tut}. 
The lines at the bottom of the figure show the time spent by a particle (a photon) 
traversing the same barrier with the speed of light.} 
\end{figure}

In fig \ref{fig:tbwd} we show the tunneling time versus the barrier width 
$d_B(F)$. The T-time shows a linear dependence with the barrier width $d_B(F)$ in 
the region $F=0.04-0.11 au$ and a limit $1/(2 I_p)$ at $d_B(F_a)=0$. 
The other limit for very large barrier width 
($F\rightarrow 0, \delta_z\rightarrow I_p$) is 
$\approx\frac{I_p}{4Z_{eff}F}=\tau_{T,sym}\approx \frac{d_B(F)}{4 Z_{eff}}$, 
which is straightforward because for very large barrier 
$\tau_{T,d}>> \tau_{T,i} \Rightarrow \tau_{T,d}\approx \tau_{T,sym}$. 
We note as seen in fig \ref{fig:tbwd} that the time spent by a particle (a photon) 
to traverse the same barrier width with the speed of light is much smaller than 
the T-time of the electron in the He-atom. 

At the limit $F=F_a$ of the sub-atomic field strength the 
tunneling process is out and an ionization process called  the 
''above the barrier decay`` is beginning. 
For supper-atomic field strength $F>F_a$, $\delta_z$ becomes imaginary (and so 
the crossing points, compare eq (\ref{xepm}), but still a real 
$x_m=\sqrt{Z{eff}/F}$), 
which indicates that the real part $\frac{1}{2I_p}$ of $\tau_{_{T,d}}$ or 
$\tau_{_{T,i}}$, is the limit for ''real''  time tunnel-ionization process.  
Indeed in this case the atomic potential is heavily disturbed and the imaginary 
part of the time $\tau_{_{T,d}}$ is then due to the release of or escaping the 
electron (at $x_m(F)$) from a lower energy level than $-Ip$ (and possibly escaping 
with a high velocity), where the ionization happens mainly by a shack-off step 
\cite{Delone:2000} (chap. 9).  
Here we see the clear difference between the quantum mechanical and the classical 
clocks \cite{Aharonov:2000,Aharonov:1961}. 
Classically we can make the interaction time with the system arbitrarily small,  
the real part of the time can be made arbitrary small, and an imaginary part 
is absent.
In quantum mechanical the tunneling-ionization time has a real part 
limit $\tau_{_{T,d}}=1/(2I_p)$, an imaginary part arise when the field strength 
is larger than the atomic field strength $F_a$, in both  terms $\tau_{_{T,i}}$ 
and $\tau_{_{T,d}}$. 
 
However in our treatment, although, $\tau_{_{T,i}}, \tau_{_{T,d}}$ have 
an imaginary part, we get a real total or symmetrical T-time 
$\tau_{{_T,sym}}=\frac{I_p}{4 Z_{eff} F}$ for ionization processes with an 
arbitrary field strength. It becomes very small for very large field strength, 
and probably it loses its validity in this regime suspecting a break of 
some symmetry and  non-linear effects arise, 
and the interaction becomes physically a different character. 
It certainly also loses its validity in the multiphoton regime, 
i.e. for large Keldysh parameter $\gamma>>1$, where $F<<F_a$. 
It is apparent from $\tau_{{_T,sym}}$, eqs (\ref{Tsym}), (\ref{Tsym1}) that 
the T-time has no imaginary part when the symmetry of the time is considered, 
i.e. when assuming  the maximally symmetrical (or quasi-self-adjoint) 
property discussed in details by Olkhovsky \cite{Olkhovsky:2009}. 
It is now the question to which extend the above relation $\frac{I_p}{4 Z_{eff} F}$, 
preserves its validity for $F>F_a$ 
(or for small $F<<F_a$, where $\gamma_K<<1$), 
where or what  is/are the limit(s) of its validity?  
Probably a break of some symmetry for  $F\ge F_a$ 
(or $F<<F_a$) can give a hint to answer this question. 
Finally we mention that for $F>F_a$ or intensities $I>I_a$ Stark-shift, 
relativistic and non-linear effects become large, the perturbation theory 
breaks down (which is valid for small parameter $F/F_a$ \cite{Delone:2000}) 
and several regions appear at intensities larger than the appearance intensity 
$I_a$, such as the critical  $I_c$ and the saturation $I_s$ intensities, 
$I_s>I_c>I_a$,  where  multiple ionization occurs \cite{Delone:2000} (chap. 7, 9).
\subparagraph{Conclusion}\label{sec:con}
We presented in this work an analysis for the tunneling process and the tunneling 
time in attosecond experiment and found an accurate and simple relation to 
calculate the tunneling time for the important case of He-atom, where a reliable  
experimental data are available. 
Our result (especially the T-time $\tau_{_{T,d}}$) was shown to be in excellent 
agreement with the experiment \cite{Eckle:2008s,Eckle:2008} and with the Feynman 
path integral treatment of \cite{Landsman:2014II} although for small field strength 
our result of $Z_{eff,K}$ tends to agree better with experimental values.
Note in figs \ref{fig:tut}-\ref{fig:tbwd} we use for the evaluation of our 
result the same values of the field  strengths used  by the experiment, 
i.e the filed strength at maximum, see \cite{Landsman:2014II}.
The T-time in our treatment is dynamical or intrinsic-time type and represents 
a quantum clock, i.e. to observe the time form within the system and consider 
the quantum nature of the (bound) particle, 
in contrast to the classical Keldysh time which is external or parametric, 
where we indicated (one or two of) its failures to treat the T-time 
in our (study) case.
 
Further we suggest a model of a shutter to the tunneling process in attosecond 
experiment, and we think the experiment together with our tunneling
model (subsec. \ref{ssec:skm}, \ref{ssec:tunt}) offer a realization of the 
Bohr-Einstein's {\it photon box Gedanken experiment}, with the electron as 
a particle instead of the photon and with the uncertainty being determined 
from the (Coulomb) atomic potential instead of the gravitational potential. 
Our treatment suggests that a symmetry (maximally symmetrical or 
quasi-self-adjoint) \cite{Olkhovsky:2009} assumption to calculate the T-time 
is important and gives a hint to the search for a time operator in the tunneling 
process and maybe for a general time operator in quantum mechanics. 
Our result uses two  models of the effective charge $Z_{eff}$ of the left 
core $He^{+}$ that the tunneling electron experiences. 
The $Z_{_{eff,K}}=1.375$ of Kullie \cite{Kullie:1997}, that based on a similar model 
to the single-active-electron,  is better for small field strength $F\lesssim0.055$,  
whereas  $Z_{_{eff,C}}=1.6875$ of Clementi et al \cite{Clementi:1963} is more 
reliable for larger field strength, because it is based on the Hartree-Fock 
calculation, and that is justified, when the multielectron effects are 
not negligible. 

\begin{thebibliography}{10}%
\makeatletter
\providecommand \@ifxundefined [1]{%
 \ifx #1\undefined \expandafter \@firstoftwo
 \else \expandafter \@secondoftwo
\fi
}%
\providecommand \@ifnum [1]{%
 \ifnum #1\expandafter \@firstoftwo
 \else \expandafter \@secondoftwo
\fi
}%
\providecommand \enquote [1]{``#1''}%
\providecommand \bibnamefont  [1]{#1}%
\providecommand \bibfnamefont [1]{#1}%
\providecommand \citenamefont [1]{#1}%
\providecommand\href[0]{\@sanitize\@href}%
\providecommand\@href[1]{\endgroup\@@startlink{#1}\endgroup\@@href}%
\providecommand\@@href[1]{#1\@@endlink}%
\providecommand \@sanitize [0]{\begingroup\catcode`\&12\catcode`\#12\relax}%
\@ifxundefined \pdfoutput {\@firstoftwo}{%
 \@ifnum{\z@=\pdfoutput}{\@firstoftwo}{\@secondoftwo}%
}{%
 \providecommand\@@startlink[1]{\leavevmode}%
 \providecommand\@@endlink[0]{}%
}{%
 \providecommand\@@startlink[1]{%
  \leavevmode
  \pdfstartlink
   attr{/Border[0 0 1 ]/H/I/C[0 1 1]}%
   user{/Subtype/Link/A<</Type/Action/S/URI/URI(#1)>>}%
  \relax
 }%
 \providecommand\@@endlink[0]{\pdfendlink}%
}%
\providecommand \url  [0]{\begingroup\@sanitize \@url }%
\providecommand \@url [1]{\endgroup\@href {#1}{\urlprefix}}%
\providecommand \urlprefix [0]{URL }%
\providecommand \Eprint[0]{\href }%
\@ifxundefined \urlstyle {%
  \providecommand \doi [1]{doi:\discretionary{}{}{}#1}%
}{%
  \providecommand \doi [0]{doi:\discretionary{}{}{}\begingroup
  \urlstyle{rm}\Url }%
}%
\providecommand \doibase [0]{http://dx.doi.org/}%
\providecommand \Doi[1]{\href{\doibase#1}}%
\providecommand \bibAnnote [3]{%
  \BibitemShut{#1}%
  \begin{quotation}\noindent
    \textsc{Key:}\ #2\\\textsc{Annotation:}\ #3%
  \end{quotation}%
}%
\providecommand \bibAnnoteFile [2]{%
  \IfFileExists{#2}{\bibAnnote {#1} {#2} {\input{#2}}}{}%
}%
\providecommand \typeout [0]{\immediate \write \m@ne }%
\providecommand \selectlanguage [0]{\@gobble}%
\providecommand \bibinfo [0]{\@secondoftwo}%
\providecommand \bibfield [0]{\@secondoftwo}%
\providecommand \translation [1]{[#1]}%
\providecommand \BibitemOpen[0]{}%
\providecommand \bibitemStop [0]{}%
\providecommand \bibitemNoStop [0]{.\EOS\space}%
\providecommand \EOS [0]{\spacefactor3000\relax}%
\providecommand \BibitemShut [1]{\csname bibitem#1\endcsname}%
\bibitem{Eckle:2008s}%
  \BibitemOpen
  \bibfield{author}{%
  \bibinfo {author} {\bibnamefont{\scriptsize P.~Eckle}}, \bibinfo {author}
  {\bibfnamefont{A.~N.}\ \bibnamefont{Pfeiffer}}, \bibinfo {author}
  {\bibfnamefont{C.}~\bibnamefont{Cirelli}}, \bibinfo {author}
  {\bibfnamefont{A.}~\bibnamefont{Staudte}}, \bibinfo {author}
  {\bibfnamefont{R.}~\bibnamefont{D\"orner}}, \bibinfo {author}
  {\bibfnamefont{H.~G.}\ \bibnamefont{Muller}}, \bibinfo {author}
  {\bibfnamefont{M.}~\bibnamefont{B\"uttiker}},\ and\ \bibinfo {author}
  {\bibfnamefont{U.}~\bibnamefont{Keller}},\ }%
  \bibfield{journal}{%
  \bibinfo {journal} {Sience}\ }%
  \textbf{\bibinfo {volume} {322}},\ \bibinfo {pages} {1525} (\bibinfo {year}
  {2008})%
  \bibAnnoteFile{NoStop}{Eckle:2008s}%
\bibitem{Eckle:2008}%
  \BibitemOpen
  \bibfield{author}{%
  \bibinfo {author} {\bibfnamefont{P.}~\bibnamefont{Eckle}}, \bibinfo {author}
  {\bibfnamefont{M.}~\bibnamefont{Smolarski}}, \bibinfo {author}
  {\bibfnamefont{F.}~\bibnamefont{Schlup}}, \bibinfo {author}
  {\bibfnamefont{J.}~\bibnamefont{Biegert}}, \bibinfo {author}
  {\bibfnamefont{A.}~\bibnamefont{Staudte}}, \bibinfo {author}
  {\bibfnamefont{M.}~\bibnamefont{Sch\"offler}}, \bibinfo {author}
  {\bibfnamefont{H.}~\bibnamefont{Muller}}, \bibinfo {author}
  {\bibfnamefont{R.}~\bibnamefont{D\"orner}},\ and\ \bibinfo {author}
  {\bibfnamefont{U.}~\bibnamefont{Keller}},\ }%
  \bibfield{journal}{%
  \bibinfo {journal} {Nat. phys.}\ }%
  \textbf{\bibinfo {volume} {4}},\ \bibinfo {pages} {565} (\bibinfo {year}
  {2008})%
  \bibAnnoteFile{NoStop}{Eckle:2008}%
\bibitem{Muga:2008}%
  \BibitemOpen
  \bibfield{author}{%
  in\ \emph{\bibinfo {booktitle} {Time in Quantum Mechanics, Lecture Notes in
  Physics 734}},\ }%
  Vol.~\bibinfo {volume} {I},\ \bibinfo {editor} {edited by\ \bibinfo {editor}
  {\bibfnamefont{G.}~\bibnamefont{Muga}}, \bibinfo {editor}
  {\bibfnamefont{R.~S.}\ \bibnamefont{Mayato}},\ and\ \bibinfo {editor}
  {\bibfnamefont{I.}~\bibnamefont{Egusquiza}}}\ (\bibinfo {publisher}
  {Springer-Verlag Berlin},\ \bibinfo {year} {2008})%
  \bibAnnoteFile{NoStop}{Muga:2008}%
\bibitem{Hilgevoord:2002}%
  \BibitemOpen
  \bibfield{author}{%
  \bibinfo {author} {\bibfnamefont{J.}~\bibnamefont{Hilgevoord}},\ }%
  \bibfield{journal}{%
  \bibinfo {journal} {Am. J. of Phys.}\ }%
  \textbf{\bibinfo {volume} {70}},\ \bibinfo {pages} {301} (\bibinfo {year}
  {2002})%
  \bibAnnoteFile{NoStop}{Hilgevoord:2002}%
\bibitem{Aharonov:2000}%
  \BibitemOpen
  \bibfield{author}{%
  \bibinfo {author} {\bibfnamefont{Y.}~\bibnamefont{Aharonov}}\ and\ \bibinfo
  {author} {\bibfnamefont{B.}~\bibnamefont{Reznik}},\ }%
  \bibfield{journal}{%
  \bibinfo {journal} {Phys. Rev. Lett.}\ }%
  \textbf{\bibinfo {volume} {84}},\ \bibinfo {pages} {1368} (\bibinfo {year}
  {2000})%
  \bibAnnoteFile{NoStop}{Aharonov:2000}%
\bibitem{Busch:1990I}%
  \BibitemOpen
  \bibfield{author}{%
  \bibinfo {author} {\bibfnamefont{P.}~\bibnamefont{Busch}},\ }%
  \bibfield{journal}{%
  \bibinfo {journal} {Found. Phys.}\ }%
  \textbf{\bibinfo {volume} {20}},\ \bibinfo {pages} {1} (\bibinfo {year}
  {1990})%
  \bibAnnoteFile{NoStop}{Busch:1990I}%
\bibitem{Busch:1990II}%
  \BibitemOpen
  \bibfield{author}{%
  \bibinfo {author} {\bibfnamefont{P.}~\bibnamefont{Busch}},\ }%
  \bibfield{journal}{%
  \bibinfo {journal} {Found. Phys.}\ }%
  \textbf{\bibinfo {volume} {20}},\ \bibinfo {pages} {33} (\bibinfo {year}
  {1990})%
  \bibAnnoteFile{NoStop}{Busch:1990II}%
\bibitem{Cooke:1980}%
  \BibitemOpen
  \bibfield{author}{%
  \bibinfo {author} {\bibfnamefont{L.~F.}\ \bibnamefont{Cooke}},\ }%
  \bibfield{journal}{%
  \bibinfo {journal} {Am. J. of Phys.}\ }%
  \textbf{\bibinfo {volume} {48}},\ \bibinfo {pages} {142} (\bibinfo {year}
  {1980})%
  \bibAnnoteFile{NoStop}{Cooke:1980}%
\bibitem{Hauge:1989}%
  \BibitemOpen
  \bibfield{author}{%
  \bibinfo {author} {\bibfnamefont{E.~H.}\ \bibnamefont{Huge}}\ and\ \bibinfo
  {author} {\bibnamefont{St{\o}vneng}},\ }%
  \bibfield{journal}{%
  \bibinfo {journal} {Rev. Mod. Phys.}\ }%
  \textbf{\bibinfo {volume} {61}},\ \bibinfo {pages} {917} (\bibinfo {year}
  {1989})%
  \bibAnnoteFile{NoStop}{Hauge:1989}%
\bibitem{Landauer:1994}%
  \BibitemOpen
  \bibfield{author}{%
  \bibinfo {author} {\bibfnamefont{R.}~\bibnamefont{Landauer}}\ and\ \bibinfo
  {author} {\bibfnamefont{T.}~\bibnamefont{Martin}},\ }%
  \bibfield{journal}{%
  \bibinfo {journal} {Rev. Mod. Phys.}\ }%
  \textbf{\bibinfo {volume} {66}},\ \bibinfo {pages} {217} (\bibinfo {year}
  {1994})%
  \bibAnnoteFile{NoStop}{Landauer:1994}%
\bibitem{Yamada:2004}%
  \BibitemOpen
  \bibfield{author}{%
  \bibinfo {author} {\bibfnamefont{N.}~\bibnamefont{Yamada}},\ }%
  \bibfield{journal}{%
  \bibinfo {journal} {Phys. Rev. Lett.}\ }%
  \textbf{\bibinfo {volume} {93}},\ \bibinfo {pages} {170401} (\bibinfo {year}
  {2004})%
  \bibAnnoteFile{NoStop}{Yamada:2004}%
\bibitem{Sokolovski:1994}%
  \BibitemOpen
  \bibfield{author}{%
  \bibinfo {author} {\bibfnamefont{D.}~\bibnamefont{Sokolovski}}, \bibinfo
  {author} {\bibfnamefont{S.}~\bibnamefont{Brouard}},\ and\ \bibinfo {author}
  {\bibfnamefont{J.~N.~L.}\ \bibnamefont{Connor}},\ }%
  \bibfield{journal}{%
  \bibinfo {journal} {Phys. Rev. A}\ }%
  \textbf{\bibinfo {volume} {50}},\ \bibinfo {pages} {1240} (\bibinfo {year}
  {1994})%
  \bibAnnoteFile{NoStop}{Sokolovski:1994}%
\bibitem{Sokolovski:1990}%
  \BibitemOpen
  \bibfield{author}{%
  \bibinfo {author} {\bibfnamefont{D.}~\bibnamefont{Sokolovski}}, \bibinfo
  {author} {\bibfnamefont{S.}~\bibnamefont{Brouard}},\ and\ \bibinfo {author}
  {\bibfnamefont{J.~N.~L.}\ \bibnamefont{Connor}},\ }%
  \bibfield{journal}{%
  \bibinfo {journal} {Phys. Rev. A}\ }%
  \textbf{\bibinfo {volume} {42}},\ \bibinfo {pages} {6512} (\bibinfo {year}
  {1990})%
  \bibAnnoteFile{NoStop}{Sokolovski:1990}%
\bibitem{Augst:1989}%
  \BibitemOpen
  \bibfield{author}{%
  \bibinfo {author} {\bibfnamefont{S.}~\bibnamefont{Augst}}, \bibinfo {author}
  {\bibfnamefont{D.}~\bibnamefont{Strickland}}, \bibinfo {author}
  {\bibfnamefont{D.~D.}\ \bibnamefont{Meyerhofer}}, \bibinfo {author}
  {\bibfnamefont{S.~L.}\ \bibnamefont{Chin}},\ and\ \bibinfo {author}
  {\bibfnamefont{J.~H.}\ \bibnamefont{Eberly}},\ }%
  \bibfield{journal}{%
  \bibinfo {journal} {Phys. Rev. Lett.}\ }%
  \textbf{\bibinfo {volume} {63}},\ \bibinfo {pages} {2212} (\bibinfo {year}
  {1989})%
  \bibAnnoteFile{NoStop}{Augst:1989}%
\bibitem{Augst:1991}%
  \BibitemOpen
  \bibfield{author}{%
  \bibinfo {author} {\bibfnamefont{S.}~\bibnamefont{Augst}}, \bibinfo {author}
  {\bibfnamefont{D.~D.}\ \bibnamefont{Meyerhofer}}, \bibinfo {author}
  {\bibfnamefont{D.}~\bibnamefont{Strickland}},\ and\ \bibinfo {author}
  {\bibfnamefont{S.~L.}\ \bibnamefont{Chin}},\ }%
  \bibfield{journal}{%
  \bibinfo {journal} {J. Opt. Soc. Am. B}\ }%
  \textbf{\bibinfo {volume} {8}},\ \bibinfo {pages} {858} (\bibinfo {year}
  {1991})%
  \bibAnnoteFile{NoStop}{Augst:1991}%
\bibitem{Muller:1999}%
  \BibitemOpen
  \bibfield{author}{%
  \bibinfo {author} {\bibfnamefont{H.}~\bibnamefont{Muller}},\ }%
  \bibfield{journal}{%
  \bibinfo {journal} {Phys. Rev. A}\ }%
  \textbf{\bibinfo {volume} {60}},\ \bibinfo {pages} {1341} (\bibinfo {year}
  {1999})%
  \bibAnnoteFile{NoStop}{Muller:1999}%
\bibitem{Tong:2005}%
  \BibitemOpen
  \bibfield{author}{%
  \bibinfo {author} {\bibfnamefont{X.}~\bibnamefont{Tong}}\ and\ \bibinfo
  {author} {\bibfnamefont{C.}~\bibnamefont{Lin}},\ }%
  \bibfield{journal}{%
  \bibinfo {journal} {J. Phys. B}\ }%
  \textbf{\bibinfo {volume} {38}},\ \bibinfo {pages} {593} (\bibinfo {year}
  {2005})%
  \bibAnnoteFile{NoStop}{Tong:2005}%
\bibitem{Schlueter:1987}%
  \BibitemOpen
  \bibfield{author}{%
  \bibinfo {author} {\bibfnamefont{D.}~\bibnamefont{Schl\"uter}},\ }%
  \bibfield{journal}{%
  \bibinfo {journal} {Z. Phys. D}\ }%
  \textbf{\bibinfo {volume} {6}},\ \bibinfo {pages} {249} (\bibinfo {year}
  {1987})%
  \bibAnnoteFile{NoStop}{Schlueter:1987}%
\bibitem{Lange:1992}%
  \BibitemOpen
  \bibfield{author}{%
  \bibinfo {author} {\bibfnamefont{R.}~\bibnamefont{Lange}}\ and\ \bibinfo
  {author} {\bibfnamefont{D.}~\bibnamefont{Schl\"uter}},\ }%
  \bibfield{journal}{%
  \bibinfo {journal} {J. Quant. Spectrosc. Radiat. Transfer}\ }%
  \textbf{\bibinfo {volume} {48}},\ \bibinfo {pages} {153} (\bibinfo {year}
  {1992})%
  \bibAnnoteFile{NoStop}{Lange:1992}%
\bibitem{Dreissigacker:2013}%
  \BibitemOpen
  \bibfield{author}{%
  \bibinfo {author} {\bibfnamefont{I.}~\bibnamefont{Dreissigacker}}\ and\
  \bibinfo {author} {\bibfnamefont{M.}~\bibnamefont{Lein}},\ }%
  \bibfield{journal}{%
  \bibinfo {journal} {Chem. Phys.}\ }%
  \textbf{\bibinfo {volume} {414}},\ \bibinfo {pages} {69} (\bibinfo {year}
  {2013})%
  \bibAnnoteFile{NoStop}{Dreissigacker:2013}%
\bibitem{Delone:1998}%
  \BibitemOpen
  \bibfield{author}{%
  \bibinfo {author} {\bibfnamefont{N.~B.}\ \bibnamefont{Delone}}\ and\ \bibinfo
  {author} {\bibfnamefont{V.~P.}\ \bibnamefont{Krainov}}\ }%
  \textbf{\bibinfo {volume} {41}},\ \bibinfo {pages} {469} (\bibinfo {year}
  {1998})%
  \bibAnnoteFile{NoStop}{Delone:1998}%
\bibitem{Klaiber:2013}%
  \BibitemOpen
  \bibfield{author}{%
  \bibinfo {author} {\bibfnamefont{M.}~\bibnamefont{Klaiber}}, \bibinfo
  {author} {\bibfnamefont{E.}~\bibnamefont{Yakaboylu}}, \bibinfo {author}
  {\bibfnamefont{H.}~\bibnamefont{Bauke}}, \bibinfo {author}
  {\bibfnamefont{K.~Z.}\ \bibnamefont{Hatsagortsyan}},\ and\ \bibinfo {author}
  {\bibfnamefont{C.~H.}\ \bibnamefont{Keitel}},\ }%
  \bibfield{journal}{%
  \bibinfo {journal} {Phys. Rev. Lett.}\ }%
  \textbf{\bibinfo {volume} {110}},\ \bibinfo {pages} {153004} (\bibinfo {year}
  {2013})%
  \bibAnnoteFile{NoStop}{Klaiber:2013}%
\bibitem{Yakaboylu:2013}%
  \BibitemOpen
  \bibfield{author}{%
  \bibinfo {author} {\bibfnamefont{E.}~\bibnamefont{Yakaboylu}}, \bibinfo
  {author} {\bibfnamefont{M.}~\bibnamefont{Klaiber}}, \bibinfo {author}
  {\bibfnamefont{H.}~\bibnamefont{Bauke}}, \bibinfo {author}
  {\bibfnamefont{K.~Z.}\ \bibnamefont{Hatsagortsyan}},\ and\ \bibinfo {author}
  {\bibfnamefont{C.~H.}\ \bibnamefont{Keitel}},\ }%
  \bibfield{journal}{%
  \bibinfo {journal} {Phys. Rev. A}\ }%
  \textbf{\bibinfo {volume} {88}},\ \bibinfo {pages} {063421} (\bibinfo {year}
  {2013})%
  \bibAnnoteFile{NoStop}{Yakaboylu:2013}%
\bibitem{Keldysh:1964}%
  \BibitemOpen
  \bibfield{author}{%
  \bibinfo {author} {\bibfnamefont{L.~V.}\ \bibnamefont{Keldysh}},\ }%
  \bibfield{journal}{%
  \bibinfo {journal} {Zh. eksp. teor. Fiz.}\ }%
  \textbf{\bibinfo {volume} {47}},\ \bibinfo {pages} {1945} (\bibinfo {year}
  {1964}),\ \bibinfo {note} {[English translation: 1965, Soviet Phys. JETP, 20,
  1307]}%
  \bibAnnoteFile{NoStop}{Keldysh:1964}%
\bibitem{Faisal:1973}%
  \BibitemOpen
  \bibfield{author}{%
  \bibinfo {author} {\bibfnamefont{F.~H.~M.}\ \bibnamefont{Faisal}},\ }%
  \bibfield{journal}{%
  \bibinfo {journal} {J. Phys. B}\ }%
  \textbf{\bibinfo {volume} {6}} (\bibinfo {year} {1973})%
  \bibAnnoteFile{NoStop}{Faisal:1973}%
\bibitem{Reiss:1980}%
  \BibitemOpen
  \bibfield{author}{%
  \bibinfo {author} {\bibfnamefont{H.~R.}\ \bibnamefont{Reiss}},\ }%
  \bibfield{journal}{%
  \bibinfo {journal} {Phys. Rev. A}\ }%
  \textbf{\bibinfo {volume} {22}},\ \bibinfo {pages} {1782} (\bibinfo {year}
  {1980})%
  \bibAnnoteFile{NoStop}{Reiss:1980}%
\bibitem{Orlando:2014}%
  \BibitemOpen
  \bibfield{author}{%
  \bibinfo {author} {\bibfnamefont{G.}~\bibnamefont{Orlando}}, \bibinfo
  {author} {\bibfnamefont{C.~R.}\ \bibnamefont{McDonald}}, \bibinfo {author}
  {\bibfnamefont{N.~H.}\ \bibnamefont{Protik}},\ and\ \bibinfo {author}
  {\bibfnamefont{T.}~\bibnamefont{Brabec}},\ }%
  \bibfield{journal}{%
  \bibinfo {journal} {Phys. Rev. A}\ }%
  \textbf{\bibinfo {volume} {89}},\ \bibinfo {pages} {014102} (\bibinfo {year}
  {2014})%
  \bibAnnoteFile{NoStop}{Orlando:2014}%
\bibitem{Delone:2000}%
  \BibitemOpen
  \bibfield{author}{%
  \bibinfo {author} {\bibfnamefont{N.~B.}\ \bibnamefont{Delone}}\ and\ \bibinfo
  {author} {\bibfnamefont{V.~P.}\ \bibnamefont{Krainov}}\ }%
  \bibinfo {note} {$\!\!$. Multiphoton Processes in Atoms, 2-edition
  (Springer-Verlag 2000)}%
  \bibAnnoteFile{NoStop}{Delone:2000}%
\bibitem{Torlina:2015}%
  \BibitemOpen
  \bibfield{author}{%
  \bibinfo {author} {\bibnamefont{Torlina}}, \bibinfo {author}
  {\bibfnamefont{F.}~\bibnamefont{Morales}}, \bibinfo {author}
  {\bibfnamefont{J.}~\bibnamefont{Kaushal}}, \bibinfo {author}
  {\bibfnamefont{I.~I.}\ \bibnamefont{H.~Geert~Muller}}, \bibinfo {author}
  {\bibfnamefont{A.}~\bibnamefont{Kheifets}}, \bibinfo {author}
  {\bibfnamefont{A.}~\bibnamefont{Zielinski}}, \bibinfo {author}
  {\bibfnamefont{A.}~\bibnamefont{Scrinzi}}, \bibinfo {author}
  {\bibfnamefont{M.}~\bibnamefont{Ivanov}},\ and\ \bibinfo {author}
  {\bibfnamefont{O.}~\bibnamefont{Smirnova}},\ }%
  \bibfield{journal}{%
  \bibinfo {journal} {arXiv:}\ }%
  \textbf{\bibinfo {volume} {{\rm 402.5620v2}}} (\bibinfo {year} {2015})%
  \bibAnnoteFile{NoStop}{Torlina:2015}%
\bibitem{Ivanov:2005}%
  \BibitemOpen
  \bibfield{author}{%
  \bibinfo {author} {\bibfnamefont{M.~Y.}\ \bibnamefont{Ivanov}}, \bibinfo
  {author} {\bibfnamefont{M.}~\bibnamefont{Spanner}},\ and\ \bibinfo {author}
  {\bibfnamefont{O.}~\bibnamefont{Smirnova}},\ }%
  \bibfield{journal}{%
  \bibinfo {journal} {J. Mod. Opt.}\ }%
  \textbf{\bibinfo {volume} {52}},\ \bibinfo {pages} {165} (\bibinfo {year}
  {2005})%
  \bibAnnoteFile{NoStop}{Ivanov:2005}%
\bibitem{Aharonov:1961}%
  \BibitemOpen
  \bibfield{author}{%
  \bibinfo {author} {\bibfnamefont{Y.}~\bibnamefont{Aharonov}}\ and\ \bibinfo
  {author} {\bibfnamefont{D.}~\bibnamefont{Bohm}},\ }%
  \bibfield{journal}{%
  \bibinfo {journal} {Phys. Rev.}\ }%
  \textbf{\bibinfo {volume} {122}},\ \bibinfo {pages} {1649} (\bibinfo {year}
  {1961})%
  \bibAnnoteFile{NoStop}{Aharonov:1961}%
\bibitem{Olkhovsky:2009}%
  \BibitemOpen
  \bibfield{author}{%
  \bibinfo {author} {\bibfnamefont{V.~S.}\ \bibnamefont{Olkhovsky}},\ }%
  \bibfield{journal}{%
  \bibinfo {journal} {Adv. in Math. Phys.}\ }%
  \textbf{\bibinfo {volume} {2009}},\ \bibinfo {pages} {1} (\bibinfo {year}
  {2009}),\ \bibinfo {note} {article ID 859710}%
  \bibAnnoteFile{NoStop}{Olkhovsky:2009}%
\bibitem{Olkhovsky:1970}%
  \BibitemOpen
  \bibfield{author}{%
  \bibinfo {author} {\bibfnamefont{V.~S.}\ \bibnamefont{Olkhovsky}}\ and\
  \bibinfo {author} {\bibfnamefont{E.}~\bibnamefont{Recami}},\ }%
  \bibfield{journal}{%
  \bibinfo {journal} {Lettere Al Nuovo Cimento}\ }%
  \textbf{\bibinfo {volume} {4}},\ \bibinfo {pages} {1165} (\bibinfo {year}
  {1970})%
  \bibAnnoteFile{NoStop}{Olkhovsky:1970}%
\bibitem{Note1}%
  \BibitemOpen
  \bibinfo {note} {It can be chosen as an almost self-adjoint operator with
  practically almost any degree of the accuracy \cite {Olkhovsky:2009}}%
  \bibAnnoteFile{NoStop}{Note1}%
\bibitem{Note2}%
  \BibitemOpen
  \bibinfo {note} {Note that for a small field strength $F<<F_a$ the electron
  absorbs one or more photons and moves vertically on the energy scale emerging
  at exit point very close to the initial point $x_i$ on the coordinate scale}%
  \bibAnnoteFile{NoStop}{Note2}%
\bibitem{Landsman:2014II}%
  \BibitemOpen
  \bibfield{author}{%
  \bibinfo {author} {\bibfnamefont{A.}~\bibnamefont{Landsman}}, \bibinfo
  {author} {\bibfnamefont{M.}~\bibnamefont{Weger}}, \bibinfo {author}
  {\bibfnamefont{J.}~\bibnamefont{Maurer}}, \bibinfo {author}
  {\bibfnamefont{R.}~\bibnamefont{Boge}}, \bibinfo {author}
  {\bibfnamefont{A.}~\bibnamefont{Ludwig}}, \bibinfo {author}
  {\bibfnamefont{S.}~\bibnamefont{Heuser}}, \bibinfo {author}
  {\bibfnamefont{C.}~\bibnamefont{Cirelli}}, \bibinfo {author}
  {\bibfnamefont{L.}~\bibnamefont{Gallmann}},\ and\ \bibinfo {author}
  {\bibfnamefont{U.}~\bibnamefont{Keller}},\ }%
  \bibfield{journal}{%
  \bibinfo {journal} {Optica}\ }%
  \textbf{\bibinfo {volume} {1}},\ \bibinfo {pages} {343} (\bibinfo {year}
  {2014})%
  \bibAnnoteFile{NoStop}{Landsman:2014II}%
\bibitem{Kullie:1997}%
  \BibitemOpen
  \bibfield{author}{%
  \bibinfo {author} {\bibfnamefont{O.}~\bibnamefont{Kullie}}\ }%
  \bibinfo {note} {$\!\!$, diploma thesis (1997), presented at the
  Mathematics-Natural Science department, Christian-Albrecht-university of
  Kiel, Germany. And O. Kullie and D. Schl\"uter, work in preparation}%
  \bibAnnoteFile{NoStop}{Kullie:1997}%
\bibitem{Clementi:1963}%
  \BibitemOpen
  \bibfield{author}{%
  \bibinfo {author} {\bibfnamefont{E.}~\bibnamefont{Clementi}}\ and\ \bibinfo
  {author} {\bibfnamefont{D.~L.}\ \bibnamefont{Raimondi}},\ }%
  \bibfield{journal}{%
  \bibinfo {journal} {J. Chem. Phys.}\ }%
  \textbf{\bibinfo {volume} {38}},\ \bibinfo {pages} {2686} (\bibinfo {year}
  {1963})%
  \bibAnnoteFile{NoStop}{Clementi:1963}%
\bibitem{Hofmann:2013}%
  \BibitemOpen
  \bibfield{author}{%
  \bibinfo {author} {\bibfnamefont{C.}~\bibnamefont{Hofmann}}, , \bibinfo
  {author} {\bibfnamefont{A.~S.}\ \bibnamefont{Landsman}}, \bibinfo {author}
  {\bibfnamefont{C.}~\bibnamefont{Cirelli}}, \bibinfo {author}
  {\bibfnamefont{A.~N.}\ \bibnamefont{Pfeiffer}},\ and\ \bibinfo {author}
  {\bibfnamefont{U.}~\bibnamefont{Keller}},\ }%
  \bibfield{journal}{%
  \bibinfo {journal} {J. Phys. B}\ }%
  \textbf{\bibinfo {volume} {44}},\ \bibinfo {pages} {125601} (\bibinfo {year}
  {2013})%
  \bibAnnoteFile{NoStop}{Hofmann:2013}%
\bibitem{Landsman:2014I}%
  \BibitemOpen
  \bibfield{author}{%
  \bibinfo {author} {\bibfnamefont{A.}~\bibnamefont{Landsman}}\ and\ \bibinfo
  {author} {\bibfnamefont{U.}~\bibnamefont{Keller}},\ }%
  \bibfield{journal}{%
  \bibinfo {journal} {J. Phys. B}\ }%
  \textbf{\bibinfo {volume} {47}},\ \bibinfo {pages} {1} (\bibinfo {year}
  {2014})%
  \bibAnnoteFile{NoStop}{Landsman:2014I}%
\end{thebibliography}
%

\end{document}